\documentclass[aps,prc,reprint,nofootinbib]{revtex4-2}

\usepackage{graphicx}
\usepackage{dcolumn}
\usepackage{bm}
\usepackage{amsmath}

\usepackage{orcidlink} 
\usepackage{booktabs}
\usepackage{braket}
\usepackage{multirow}

\begin{document}

\title{Large-scale fission properties with BSkG3 I : fission paths and spontaneous fission half-lives}

\author{A. Sánchez-Fernández$^{1,2}$}
\author{W. Ryssens$^{1,2}$}
\author{S. Goriely$^{1,2}$}

\affiliation{$^1$Institut d'Astronomie et d'Astrophysique, Université Libre de Bruxelles, Brussels, Belgium}

\affiliation{$^2$Brussels Laboratory of the Universe - BLU-ULB, Brussels, Belgium}

\date{\today}

\begin{abstract}

We present our first large-scale analysis of nuclear fission properties based on the BSkG3 energy density functional. Building on the model's unprecedented agreement with ground state observables and fission data in the actinide region, we survey nearly 3300 nuclei with $80\leq Z\leq118$, spanning the region between the predicted proton and neutron drip lines. We calculate thousands of potential energy surfaces that explicitly account for triaxial and octupole deformations. We determine the corresponding least-action fission paths using microscopic collective inertias derived consistently from BSkG3 and calculate spontaneous fission (SF) half-lives within the semiclassical WKB approximation. Our results reveal a rich diversity of fission path topologies, extending well beyond the conventional two-humped picture commonly adopted in reaction codes, and a wide variety of nuclear shapes along these paths, highlighting the limitations of exploring only one collective degree of freedom. In particular, triaxiality affects not only inner barriers in the actinide region but also outer barriers and single barriers in several regions of the superheavy nuclei. Our predicted SF half-lives follow trends similar to those of previous large-scale studies, although significant differences emerge near the neutron drip line. The resulting dataset, including fission barriers, SF half-lives, and saddle-point nuclear level densities, is available has been incorporated into the TALYS reaction code and the BRUSLIB repository. It provides a microscopic foundation for nuclear reaction calculations and astrophysical applications, particularly rapid neutron-capture simulations, which require reliable fission properties of nuclei far from the valley of stability.

\end{abstract}

\maketitle

\section{Introduction}

The rapid neutron-capture process or $r$-process accounts for approximately half of the nuclei heavier than iron and is a primary driver of the chemical evolution of the Universe. Extreme astrophysical events such as neutron star mergers (NSMs)~\cite{Kasen2017} and, potentially,  magneto-rotational supernovae~\cite{Reichert2021} expose matter to intense neutron fluxes that push nuclear pathways far from the valley of stability into the region of very neutron-rich nuclei. In these exotic, neutron-rich nuclei, fission (spontaneous, neutron-induced or $\beta$-delayed) could compete with neutron capture and $\beta$-decay. Nuclear fission is a crucial component of $r$-process dynamics: fission processes set the upper mass limit of synthesized nuclei and the recycling of fission fragments influences the final abundance distribution~\cite{Goriely15,goriely2015d,panov2005,giuliani2020,martinez-pinedo2007}. The importance of fission extends to observational multi-messenger studies: together with $\beta$-decay, the spontaneous fission (SF) of long-lived transactinides powers kilonovae~\cite{arcavi2017,barnes2016,cote2018,zhu2018,vassh2019}. The correct interpretation of past and future multi-messenger observations of NSMs and the associated nucleosynthesis events is thus contingent on our understanding of fission processes.

Fission rates and yields are but one of many types of nuclear data that is required for simulations of these astrophysical processes~\cite{Goriely23}. The experimental determination of all required  nuclear properties for thousands of short-lived nuclei is simply not feasible, which is why nuclear theory needs to fill the gap. Reliably extrapolating any nuclear quantity toward the neutron dripline based on available experimental data is difficult, but this is particularly true for fission properties for two reasons. First, the theoretical description of fission still poses enormous challenges~\cite{Schunck16,Bender20,schunck2022}, even decades after its discovery~\cite{hahn1938} and despite its many applications in basic science and society~\cite{hofmann2000,mention2011,Goriely15}. Second, the available fission data is scarce: for instance, only 136 known SF half-lives~\cite{Sanchez2026a}, 77 empirical barriers~\cite{capote2009} and a handful of $\beta$-delayed fission probabilities~\cite{andreyev2013} are available.

Ab initio modelling of the structure of heavy nuclei - let alone their fission dynamics - is not feasible today~\cite{Hergert2020}, which means that microscopic models based on nuclear energy density functionals (EDFs) are best placed today to furnish the fission data needed. While such models are not free of adjustable parameters, they are grounded in quantum many-body theory yet remain sufficiently tractable for global studies across the nuclear chart~\cite{Bender03,Schunck19}.
This means that the extrapolations of EDF-based models for fission and other observables should be more reliable than those of more phenomenological models, whether semiclassical~\cite{Khuyag20}, microscopic-macroscopic~\cite{Moller2016,Bao13,Blanco23}, fully data-driven~\cite{yuan2024} or a machine-learning enabled combination of these~\cite{li2024b}. Although the predictive power of EDF-based models is hard to quantify~\cite{goriely2014,schunck2015a}, there is a simple argument to be made for the reliability of their predictions: EDFs manage to accurately\footnote{Unfortunately but not unexpectedly, EDF accuracy for fission properties is no match yet for highly parameterized modelling targeting individual properties of known nuclei~\cite{schunck2022,capote2009}.} and \emph{simultaneously} describe known ground state (masses, radii, \ldots) and fission properties~\cite{Goriely07,capote2009,Lemaitre2018,lemaitre2019,Grams23,Sanchez2026a} with a single parameter set, whereas simpler models tend to focus on a single observable and/or recommend separate parameterisations for different purposes~\cite{Moller2016}.

Nevertheless, delivering global fission properties with accuracy poses two challenges to EDF models. The first is computational: the microscopic nature of EDF-based models and the typical timescales of fission processes mean that a full time-dependent treatment is only feasible for isolated nuclei~\cite{bulgac2020b}. An alternative is the adiabatic approximation which treats the evolution of the nucleus from compact ground state (GS) to a scission configuration in a low-dimensional collective space~\cite{schunck2016}. In that case, the bottleneck is the construction of the collective space by means of a large number of constrained Hartree-Fock-Bogoliubov (HFB) calculations; several collective degrees of freedom are known to be relevant - the impact of several multipole moments, the neck and the particle dispersion number have been studied widely~\cite{schunck2016} - but the most sophisticated practical calculations so far considered only three dimensions and restricted themselves to a single nucleus~\cite{sadhukhan13,sadhukhan2016,matheson2019,Flynn2022}. Because of these computational reasons, only a limited number of EDF-based surveys targeting the fission properties of thousands of nuclei exist~\cite{mamdouh2001,Goriely07,Erler12,Lemaitre2018,Giuliani18,taninah2020}; all are limited to one-dimensional collective spaces. The other challenge is accuracy for known data: it is well-established that accuracy for fission properties requires that some information on the physics of elongated nuclear shapes enters the parameter adjustment of an EDF-based model~\cite{bartel1982,Goriely07,jodon2016,Kortelainen12,Ryssens23,Grams23,dacosta2024}. However, even fission (pseudo-)observables such as barriers and/or isomers are rarely incorporated in fitting protocols; of the existing large-scale surveys, only two  employed models - BSk14 and D1M - that were \emph{a priori} informed of the physics of fission~\cite{Goriely07,Lemaitre2018}. As a result, these two parameterizations significantly outperform the other models that have been applied at scale w.r.t.~the reproduction of known (empirical) barriers and fission isomer excitation energies - see Tab. 3 in Ref.~\cite{Ryssens23}.

We present here the results of our large-scale survey of fission properties with the BSkG3 EDF~\cite{Grams23}: the fission properties of nearly 3300 nuclei with $80 \le Z \le 118$ and spanning the entire range between predicted proton and neutron drip lines. This work simultaneously addresses both challenges facing EDF-based fission models. First, ours is the first large-scale survey that moves beyond one-dimensional collective spaces: by constructing two different two-dimensional collective spaces, we account for two quadrupole ($\beta_{20}, \beta_{22}$) and one octupole ($\beta_{30}$) deformation mode of the nuclear density. In this way, we consistently account for axial and reflection asymmetry. Second, the accuracy of BSkG3 for \emph{known} fission properties is unsurpassed at the moment of writing: the model reproduces the 45 primary empirical barriers of actinide nuclei in the RIPL-3 library with a root-mean-square (rms) deviation of just 0.330 MeV~\cite{Grams23} and describes the measured SF lifetimes within three orders of magnitude~\cite{Sanchez2026a,Sanchez2026b}. We emphasize that this performance for fission properties was not obtained by compromising the models accuracy for GS properties: BSkG3 also offers a competitive global description of, e.g., all the 2457 known nuclear masses above $^{16}$O with an rms deviation $\sigma = 0.631$ MeV that is competitive with other large-scale EDF models~\cite{goriely2009c,goriely2016a,scamps2021,ryssens2022,grams2025a,grams2026}.

To facilitate future applications of our work, we provide the calculated least action fission paths and SF half-lives for all nuclei. We also provide predictions for the nuclear level densities at the saddle points of the fission paths based on the combinatorial method~\cite{hilaire2001} applied to consistent BSkG3 single-particle level schemes along the lines of~\cite{goriely2026}. The results are available in the latest version of the TALYS code\cite{Koning23,TALYS_GitHub} and the BRUSLIB repository\cite{Bruslib}. While this paper focuses on fission barriers and SF, the data we provide is sufficient for Hauser-Feschbach calculations to estimate neutron-induced fission cross sections with reaction codes such as TALYS. Calculating $\beta$-delayed fission probabilities based on our results is also possible, provided one obtains predictions for the $\beta$-feeding of the daughter nuclei from some other source~\cite{bara2025a}.

This paper is organized as follows. Section~\ref{section_II} summarizes the theoretical framework for treating fission within the adiabatic approximation using BSkG3 while Sect.~\ref{section_III} describes practical aspects of our computations. We provide a detailed comparison of our results for actinide nuclei with available empirical barriers in Sect.~\ref{section_IV}. Section~\ref{section_V}  offers a more global view of our fission predictions and a (preliminary) discussion of their impact on $r$-process simulations. Sect.~\ref{section_VII} concludes with a summary and outlook.

\section{Theoretical framework}\label{section_II}
In this work, we employed the BSkG3 EDF model. While a detailed discussion of the model and its performance can be found in Ref.~\cite{Grams23}, we briefly summarize its structure here.

We represent a nuclear configuration as a single Bogoliubov state and define its total energy as:
\begin{equation}
    E_{\mathrm{tot}}=E_{\mathrm{HFB}}+E_{\mathrm{corr}},
\end{equation}
where $E_{\mathrm{HFB}}$ is the self-consistent mean-field energy and $E_{\mathrm{corr}}$ combines different corrections to account for correlations that are missing in single-reference EDF models. We write the mean-field energy as a combination of five terms:
\begin{equation}
    E_{\mathrm{HFB}} = E_{\mathrm{kin}}+E_{\mathrm{Sky}}+E_{\mathrm{pair}}+E_{\mathrm{Coul}}+E_{\mathrm{cm}}^{(1)},
\end{equation}
which are, respectively (i) the kinetic energy of the nucleons, (ii) the Skyrme EDF that models the strong interaction between nucleons in the particle-hole channel, (iii) the pairing EDF that models the strong interaction in the particle-particle channel, (iv) the Coulomb energy and (v) the one-body part of the centre-of-mass correction - see Refs.~\cite{Scamps21,Grams23} for details. In the particle-hole channel, BSkG3 relies on an extended Skyrme form that features additional density dependencies in order to reconcile an accurate modelling of nuclear masses with the existence of heavy neutron stars~\cite{chamel2009}. In the pairing channel, we employ zero-range pairing terms - with and without gradients of the matter density - and use the strategy of Ref.~\cite{chamel2010b} to obtain microscopic pairing strengths based on advanced many-body calculations of pairing gaps in infinite nuclear matter~\cite{chamel2008a}.

The correction term $E_{\mathrm{corr}}$ consists of four parts,
\begin{equation}\label{eq:e_corr}
E_{\mathrm{corr}}=E_\mathrm{W} + E_{\mathrm{rot}}+E_{\mathrm{vib}}+E_{\mathrm{cm}}^{(2)},
\end{equation}
which are, respectively, the Wigner, rotational, vibrational and the two-body part of the center of mass corrections along with the Wigner energy - see Refs.~\cite{Scamps21,ryssens2022} for more details. The latter three are all crucial to the success of BSkG3 for fission: including the two-body part of the center-of-mass correction is necessary to obtain a realistic surface tension when fitting Skyrme parameterizations to masses~\cite{dacosta2024}, while it is the
parameters of the rotational and vibrational collective corrections that provided the degrees of freedom to fit the RIPL-3 empirical barriers during the original model adjustment~\cite{ryssens2022}. More precisely, the vibrational and rotational terms are given by
\begin{equation}  
    E_{\mathrm{rot/vib}}=-\sum_{\mu=x,y,z} f_{\mu}^{\mathrm{rot/vib}} \frac{\braket{\hat{J}^2_\mu}}{2\mathcal{I}_\mu},  
  \end{equation}
where the $\mathcal{I}_{\mu}$ are the Belyaev rotational moments of inertia (MOI) obtained consistently from our Bogoliubov states and the factors $f^{\rm rot/vib}$ absorb the parameter-dependence of both collective corrections~\cite{ryssens2022}.
All energies discussed in this paper include all corrections and we emphasize that all parameters were determined once and for all during the construction of BSkG3~\cite{Grams23}.

\subsection{Potential energy surfaces}

The potential energy surface (PES) describes how the total energy of a nucleus changes as its shape evolves. By exploring the PES along various deformation parameters, one can identify fission barriers, metastable states, and the most favorable paths the nucleus follows from its GS to scission. This is achieved by modifying the single-particle Hamiltonian as follows:
\begin{equation}
\hat{h}' = \hat{h} - \sum_i \lambda_i \hat{O}_i,
\end{equation}
where $\lambda_i$ are the Lagrange multipliers associated with the constraint operators $\hat{O}_i$. In particular, we impose constraints on the nuclear shape through the mass multipole moments defined as:
\begin{equation}
\langle \hat{Q}_{\ell m} \rangle = \int d^3r\, \rho(\mathbf{r})\, r^{\ell} \text{Re} Y_{\ell m}(\theta, \phi),
\end{equation}
where $\rho(\mathbf{r})$ is the nuclear matter density and $Y_{\ell m}$ is a spherical harmonic of degree $l$ and order $m$. In what follows, we will use three of these as collective coordinates: both quadrupole moments $\ell = 2, m = 0/2$ and one octupole moment ($\ell = 3, m = 0$). These three collective degrees of freedom are essential for an accurate description of fission in the actinide region: $\langle \hat{Q}_{2 0} \rangle$ governs the overall elongation of the nucleus toward scission, while finite $\langle \hat{Q}_{22}\rangle$ characterizes triaxial deformation that can significantly reduce both inner and outer barriers for actinide nuclei~\cite{Ryssens23}. The octupole moment $\langle \hat{Q}_{30} \rangle$ introduces reflection-asymmetric shapes and plays a key role in generating the mass asymmetry of the fission fragments and further lowering the barriers.

All PESs and paths will be shown as a function of the dimensionless deformation parameters given by
\begin{equation}
\beta_{lm}=\frac{4\pi}{3R^l}\braket{\hat{Q}_{lm}},
\end{equation}
where $R=1.2A^{1/3}$ fm and $\hat{Q}_{lm}$.

One of the most common challenges in computing fission PESs is the emergence of discontinuities \cite{Dubray12,Zdeb21,Lau22}. In two-dimensional PESs, only two collective coordinates are explicitly constrained, while the remaining degrees of freedom (e.g., higher-order multipole deformations, pairing, etc.) are left to variationally find their energy-optimal values. As a result, smooth energy landscapes are not guaranteed, and abrupt transitions may arise in the unconstrained coordinates.

Based on prior experience, PESs mapped in the $(\beta_{20}, \beta_{30})$ plane generally display fewer discontinuities than those constructed in the $(\beta_{20}, \beta_{22})$ plane. This behavior can be partly attributed to the enforcement of axial symmetry ($\beta_{22}=\beta_{32}=0$), which reduces the number of active collective degrees of freedom. 
Furthermore, we decided to adopt a two-step strategy:

\begin{enumerate} 
\item Compute the minimum energy path (MEP) on the $(\beta_{20}, \beta_{30})$ surface to extract the evolution of $\beta_{30}$ as a function of $\beta_{20}$. 
\item Construct the triaxial PES in the $(\beta_{20}, \beta_{22})$ plane while constraining $\beta_{30}$ to the values obtained in the previous step. 
\end{enumerate}

In this work, we obtained our PESs with a spacing $\Delta\beta_{20}=\Delta\beta_{30}\sim 0.1$, while $\Delta\beta_{22}\sim 0.04$.

\begin{figure}[htbp] 
\begin{center}
\includegraphics[width=1.0\linewidth]{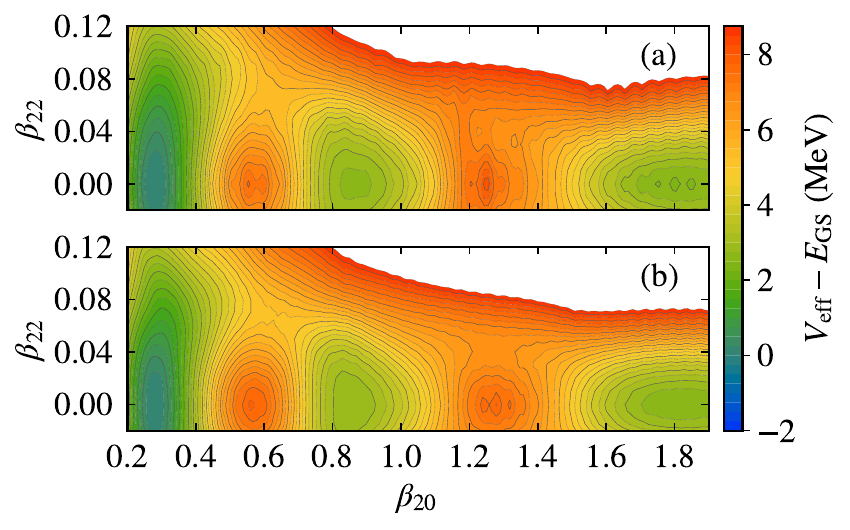} 
\end{center} 
\caption{\label{U244_CCU_CCC} $(\beta_{20},\beta_{22})$ PESs of $^{244}$U: (a) without additional constraints and (b) with $\beta_{30}$ constrained to the values of the MEP across the $(\beta_{20},\beta_{30})$ surface.} \end{figure}

This procedure can be regarded as a ``poor-man's'' approach to constructing a three-dimensional fission path. It retains the simplicity of working with two-dimensional surfaces while effectively incorporating the three most relevant deformation degrees of freedom. As an illustrative example, we consider $^{244}$U, whose $(\beta_{20}, \beta_{22})$ PES exhibits a rugged region for $\beta_{20}=[1.0, 1.5]$, as shown in Fig.~\ref{U244_CCU_CCC}(a). By constraining the octupole moment to follow the MEP on the axial $(\beta_{20}, \beta_{30})$ surface, the triaxial PES becomes significantly smoother, as seen in panel (b). This strategy was inspired by the observation that the energy-optimal $\beta_{30}$ of \emph{actinide} nuclei does not vary much  $\beta_{22}$ at fixed $\beta_{20}$~\cite{Ryssens23}. Although it is doubtful that this feature is universal, we nevertheless adopt the two-step strategy for all nuclei for three reasons: (i) it reduces the possibility for discontinuities, (ii) it incorporates more information on the third collective coordinate compared to calculations based on a single two-dimensional PES and (iii) three-dimensional PESs are beyond our computational resources.

Even though this two-step approach allows us to trace a smooth path from the GS to a semiclassical outer turning point or a low-neck configuration, it does not ensure continuity in other collective degrees of freedom, such as higher-order multipole moments. Discontinuities can still arise near scission, for instance as sudden jumps in the hexadecapole moment $\beta_{40}$ due to the crossing between the fusion and fission valleys \cite{Regnier19}. While more sophisticated methods have recently been developed to produce fully continuous PESs—by constraining state overlaps \cite{Carpentier24} or controlling the neck size \cite{Han21}—our approach prioritizes simplicity, ensuring scalability for the computation of the thousands of PESs needed to generate fission data for the r-process.

When dealing with odd nuclei--including even-odd, odd-even, and odd-odd systems--each we chose each point on the PES as the lowest-energy blocked quasiparticle configuration, obtained by using the equal filling approximation \cite{PerezMartin08}, which preserves time-reversal symmetry and thereby significantly reduces the computational cost \cite{Ryssens23,schunck2023}. For configurations that preserve parity (i.e., PES points with $\beta_{30} \approx 0$), we perform separate blocking calculations for quasiparticle states of both positive and negative parity for each nucleon species. Although the $z$-signature is preserved in all calculations, we restrict ourselves to positive signature states. While this approach is sufficiently general for odd-mass nuclei, it is less straightforward to apply to odd-odd nuclei, since states with different $\eta$ can differ slightly in energy. Accounting for this effect would require performing an even larger number of blocked calculations for odd-odd systems than are already performed. In any case, the spin-isospin terms used in modern EDFs—the terms responsible for breaking the degeneracy—are not capable of reproducing empirical observations regarding the relative alignment of the odd neutron and proton in odd-odd systems, as discussed in \cite{Robledo14}.

\subsection{Determination of the fission paths and tunneling}\label{th_fis_paths}

While a static description of fission yields the smallest possible fission barriers along the MEP, in this work we minimize the action integral
\begin{equation}\label{eq:action}
S = \int_{s_{\rm in}}^{s_{\rm out}} ds\, \sqrt{\frac{2}{\hbar^2} \mathcal{M}_{\rm eff}(s) \left( V_{\rm eff}(s) - E_{\mathrm{GS}} \right)},
\end{equation}
thereby accounting for the fission dynamics via the collective mas, $\mathcal{M}_{\rm eff}(s)$, and determining the most probable trajectory toward scission into two fragments.  
In Ref.~\cite{Sanchez2026c}, we showed that a microscopic PES can have multiple coexisting paths that locally minimize the action integral. In our previous work we referred to these as stationary action paths (SAPs). We found that the SAPs associated with SF are generally axial, whereas, with increasing excitation energy, paths that evolve through triaxial deformations can be favored in terms of the global action minimum. However, two observations motivate the use of triaxial SAPs in the present work: (i) both the axial and triaxial paths give rise to SF lifetimes that match equally well known data \cite{Sanchez2026b} and (ii) neutron-induced fission cross sections are significantly better reproduced when triaxial paths are taken into account. We therefore consider paths that locally minimize the action and traverse through triaxial fission barriers. For simplicity, and consistency with the usual terminology, we will refer to these triaxial paths as least-action paths (LAPs) hereafter.

In Eq.~\eqref{eq:action}, $s_{\rm in}$ is set as the GS, defined as the absolute minimum of the PES up to $\beta_{20} = 0.50$, and $s_{\rm out}$ corresponds to either (i) a nuclear configuration satisfying $E = E_{\rm GS}$, i.e., the semiclassical outer turning point of the tunneling process, or (ii) a low-neck configuration just prior to scission. In this context, we take $\braket{\hat{Q}_{\rm neck}} = 6$, with \cite{Schunck14}
\begin{equation}
\hat{Q}_{\rm neck} = \exp{\left[-\frac{(z-z_0)^2}{a_0^2}\right]},
\end{equation}
where $a_0 = 1$ fm, and $z_0$ is the neck position, defined as the minimum between the two density maxima of the fission pre-fragments. For each PES, we explore different possible exit points and select the configuration according to the criterion (i) or (ii) that occurs first, i.e. at the smallest value of $\beta_{20}$. Additionally, the collective mass, also referred to interchangeably as the effective inertia, is defined as

\begin{equation}
\mathcal{M}_{\rm eff}(s) = \sum_{ij} M_{ij}\frac{d\beta_i}{ds} \frac{d\beta_j}{ds},
\label{coll_mass}
\end{equation}
where $M_{ij}$ are the components of the collective inertia tensor, wih indices $i,j$ running over the explored collective coordinates. We computed the inertia tensor within the adiabatic time-dependent HFB (ATDHFB) approach in the cranking approximation \cite{Baran2011,Giuliani18}, consistent with the PESs, both derived using the same EDF model.
As these inertias were not obtained using the generator coordinate method (GCM), we set $V_{\rm eff} = E_{\rm tot}$ without the subtraction of the zero-point energy contribution \cite{Schunck16}. 

All paths considered in this work, both MEP and LAP, were determined using the PyNEB code~\cite{Flynn2022}. We use the action integral along the $(\beta_{20},\beta_{22})$ LAP to obtain the penetration probability through the fission barrier in the WKB approximation~\cite{Berry1972}:
\begin{equation}
P = \frac{1}{1 + e^{2S}},
\end{equation}
directly linked to the SF half-live as:
\begin{equation}
t_{1/2}^{\mathrm{sf}} = \frac{\ln(2)}{nP},
\end{equation}
where $n$ is the number of assaults on the fission barrier per unit time, typically assumed to be $10^{20.38}~\text{s}^{-1}$~\cite{Baran1978}.

\section{Numerical aspects} \label{section_III}

\subsection{Simulation volume and grid spacing} \label{sec_III_A}

Although the accuracy of a spatial coordinate representation has been already analyzed in detail~\cite{Ryssens15}, we revisit here some numerical aspects to find a compromise between computational efficiency and numerical accuracy that can be applied to the large-scale production of fission PESs.

The single-particle wave functions are represented on a Lagrange mesh, characterized by the number of points along each axis, $(N_x,N_y,N_z)$, and a uniform mesh spacing $dx$, common to all three axes. These parameters determine the size of the simulation volume and the spatial resolution of the representation. A compromise must therefore be found between ensuring a sufficiently large simulation volume to accommodate the fissioning nucleus and keeping the number of mesh points as small as possible to limit the computational cost. Since the length of each axis is given by $(2N-1)dx$, increasing the simulation volume requires either increasing the number of mesh points or increasing the mesh spacing, the latter reducing the spatial resolution. We first investigate the minimum axis length required to accurately describe highly elongated nuclear shapes and then examine the impact of the mesh spacing on the calculated results.

\begin{figure}[htbp]
\begin{center}
\includegraphics[width=\linewidth]{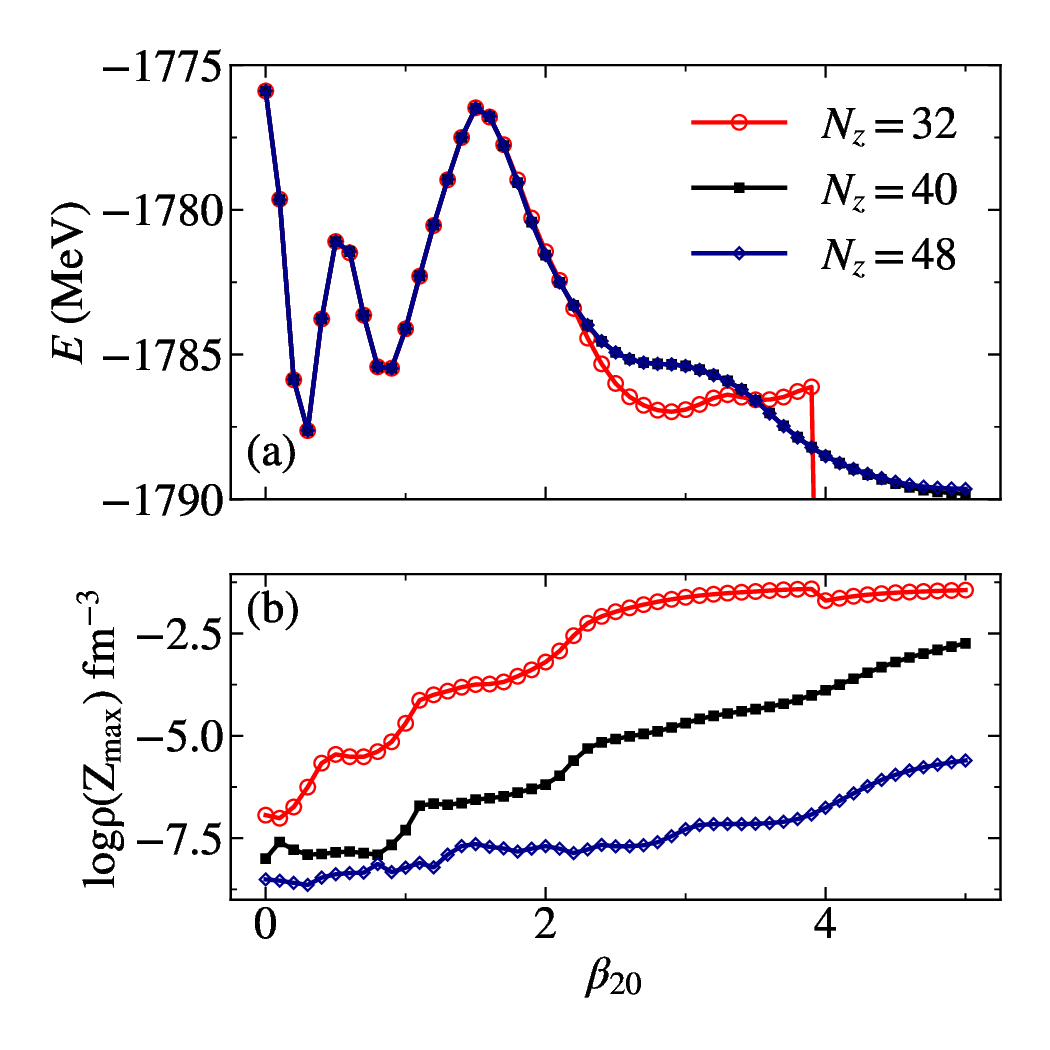}
\end{center}
\caption{\label{comparison_nz} (a) Axial fission path of $^{236}$U with three different number of points of the Lagrange mesh in the OZ-axis, fixing $dx=1.0$ fm. and (b) nuclear matter density at the edge of the simulation box.}
\end{figure}

In Fig.~\ref{comparison_nz}(a), the axial, mass-symmetric fission path\footnote{We denote the energy along the one-dimensional path as $E$, reserving $V_{\rm eff}$ for the two-dimensional energy surface.} for $^{236}$U is shown for different lengths of the $z$-axis, defined by varying $N_z$ while keeping $dx=1.0$~fm. We observe that all three cases yield essentially identical absolute energies up to $\beta_{20}=2.25$, just beyond the outer saddle. Beyond this point, however, the results obtained with the shortest axis deviate significantly. This deviation arises because the two pre-fragments are pushed against the boundaries, producing artificial deformations that culminate in an early split near $\beta_{20}\sim 4.0$. Indeed, in Fig.~\ref{comparison_nz}(b) we plot the logarithm of the nuclear matter density at the edges of the OZ-axis, showing an increase above $10^{-3}$ fm$^{-3}$ after the second saddle, confirming that the nucleus is contacting the walls. Based on this analysis, we ensured that the particle density remains below $10^{-3}$ fm$^{-3}$ near the edges of the simulation box for all intermediate configurations between the GS and the scission point. In practice, setting $N_z=40$ was sufficient to accurately describe all deformations for the nuclei considered in our large-scale survey.

\begin{figure}[htbp]
\begin{center}
\includegraphics[width=\linewidth]{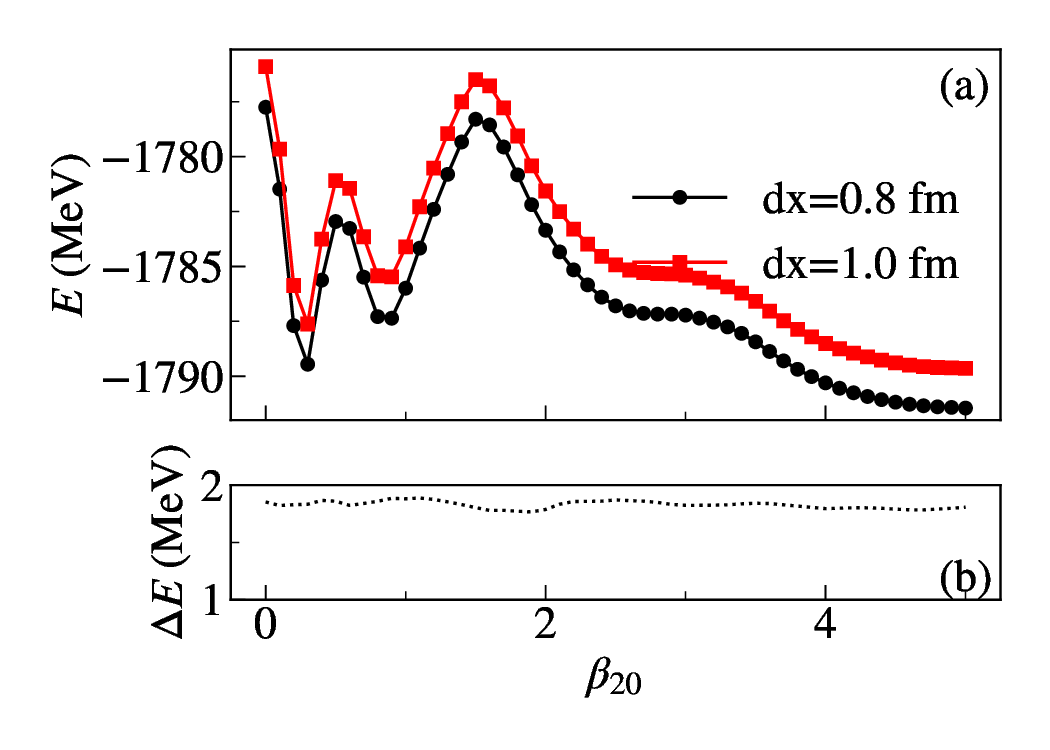}
\end{center}
\caption{\label{comp_box} Same as Fig.~\ref{comparison_nz}, but (a) comparing energy curves for different grid spacings while keeping the simulation volume fixed. (b) Energy difference between both sets of results.}
\end{figure}

To benchmark the accuracy as a function of grid spacing, we fixed the same simulation volume, while using spacings of $dx=0.8$ fm and $dx=1.0$ fm. Specifically, we performed two sets of calculations: $N_x=N_y=32$, $N_z=50$ with $dx=0.8$~fm, and $N_x=N_y=24$, $N_z=40$ with $dx=1.0$~fm. In Fig.~\ref{comp_box}(a), both energy curves are shown, while panel (b) presents the difference between them.

We found that both sets of parameters yield essentially the same energy curve, with the finer mesh producing absolute energies about 2 MeV lower than the coarser one. While this difference is relevant for absolute energies, fission properties are primarily determined by relative energies. As Fig.~\ref{comp_box}(b) shows, the energy differences remain nearly constant across the entire range of deformations, with only minor variations of around 100~keV. This demonstrates that a coarser grid is sufficient, allowing us to reduce the number of points on the Lagrange mesh and thus lower the computational cost. Consequently, for the large-scale production calculations, we fixed $N_x=N_y=24$, $N_z=40$ and $dx= 1.0$ fm.

\subsection{Number of single-particle wave functions}

In addition to the Lagrange mesh parameters, the number of single-particle wave functions also has a significant impact on both numerical accuracy and computational cost. In the construction of PESs, the rotational and vibrational MOI contribute to the total nuclear energy through the correction terms given in Eq.~\eqref{eq:e_corr}. Moreover, the collective MOI are a key ingredients in the calculation of the effective inertia along the fission path. Within the ATDHFB approach, all particle and hole states contribute to the MOI \cite{Baran2011,Giuliani18}, making them sensitive to the number of single-particle wave functions, $n_{\mathrm{wf}}$. Even though we preserved time-reversal symmetry, allowing us to compute only one state of the Kramers-degenerate pair \cite{Ryssens15b}, increasing $n_{\mathrm{wf}}$ substantially raises the memory requirements and CPU time needed for a single HFB calculation, leading to a direct trade-off between numerical accuracy and computational efficiency. 

To investigate the impact of the $n_{\mathrm{wf}}$, we selected $^{332}$Hs ($Z=108$, $N=224$), which belongs to the top 10\% of the heaviest nuclei relevant for astrophysical applications. In particular, we computed the LAP across the $(\beta_{20},\beta_{22})$ PES, varying the number of wave functions included. 

\begin{figure}[htbp]
\begin{center}
\includegraphics[width=\linewidth]{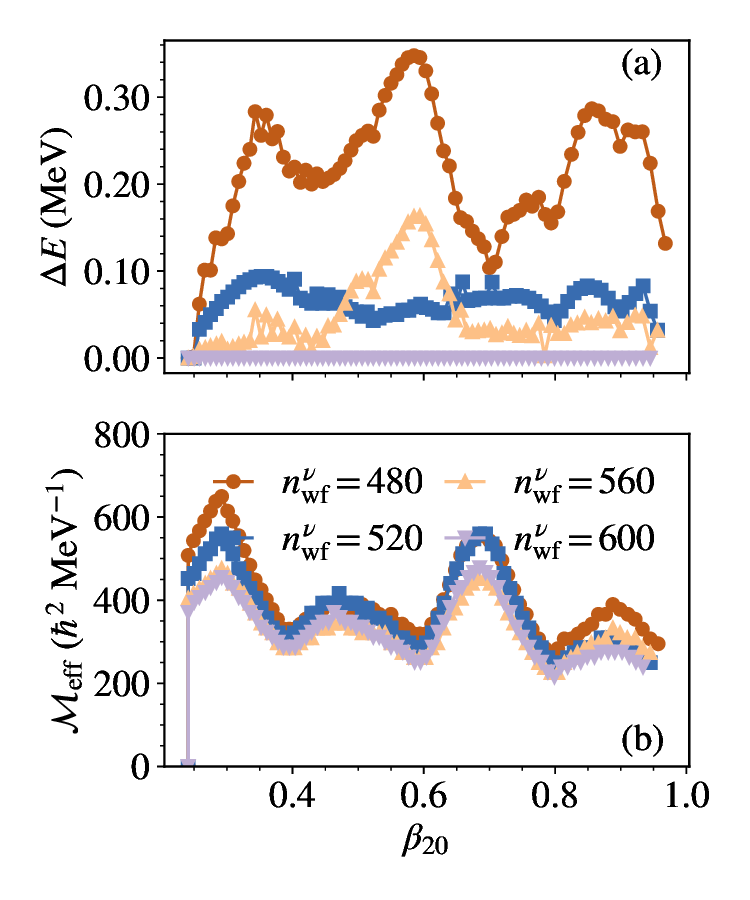}
\end{center}
\caption{\label{diff_spwf} (a) 1-dimensional LAP obtained on the $(\beta_{20},\beta_{22})$ PES of $^{332}$Hs for different number of proton and neutron s.p. wave functions and (b) the collective mas.}
\end{figure}

In Fig.~\ref{diff_spwf}(a), we show the energy differences along the LAP calculated for various values of $n_{\mathrm{wf}}^{\nu}$, relative to the case with the largest number of neutron wave functions, $n_{\mathrm{wf}}^{\nu}=600$, while keeping the number of proton wave functions fixed at $n_{\mathrm{wf}}^{\pi}=380$. The energy differences along the fission paths are generally on the order of a few hundred keV, with notable deviations near $\beta_{20}=0.6$ and 0.9, corresponding to the two local maxima of the path. Calculations with more than 520 neutron single-particle wave functions yield differences below 100 keV, with occasional exceptions.

For the GS, we found that the highest quasiparticle energies for $n_{\mathrm{wf}}^{\nu}=480$ and 600 are approximately 40.1 MeV, corresponding to an effective 40 MeV cutoff in a harmonic-oscillator basis-expansion approach.

Fig.\ref{diff_spwf}(b) shows the effective inertia along the LAP. Using fewer neutron wave functions leads to more pronounced peaks, whereas in the reference case with 600 neutron wave functions, the behavior is smoother. The differences between 560 and 600 wave functions for both energy and effective inertia are minimal.

\begin{figure}[htbp]
\begin{center}
\includegraphics[width=\linewidth]{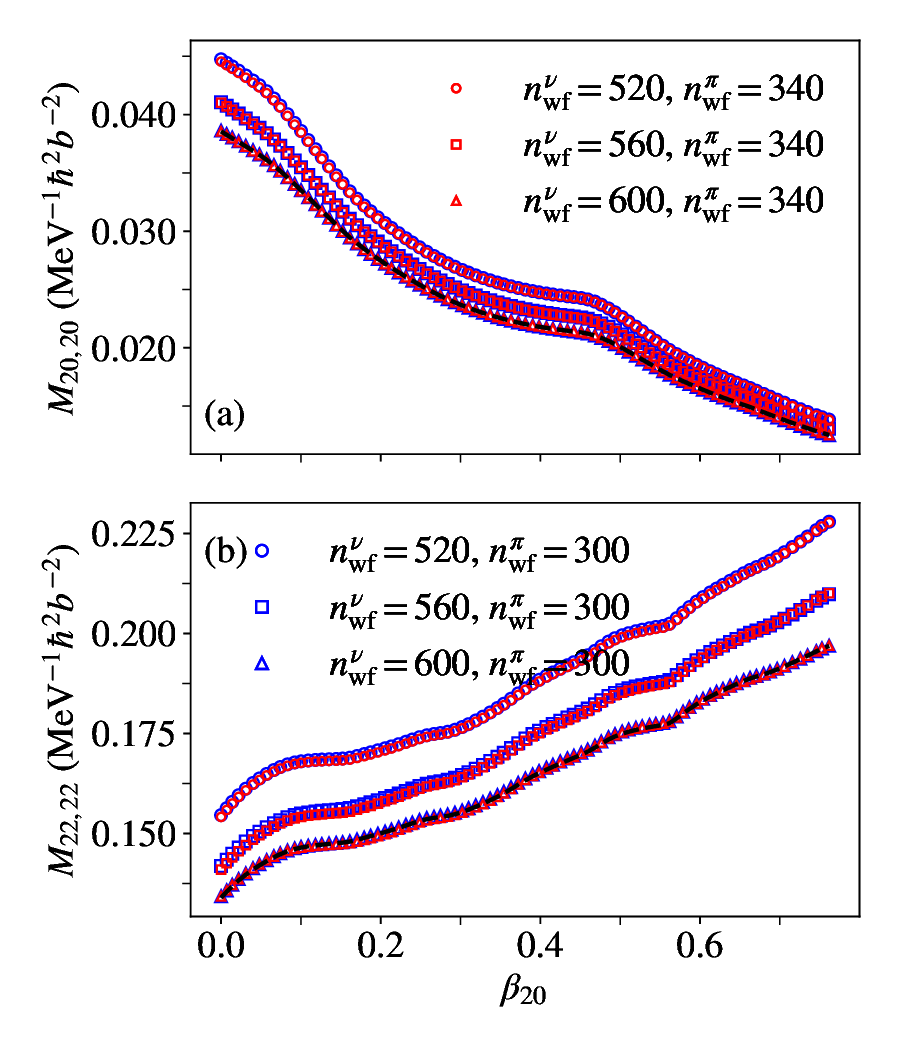}
\end{center}
\caption{\label{M_tensor} Diagonal components of the inertia tensor in the ATDHFB approximation for the collective coordinates (a) $\braket{\hat{Q}_{20}}$ and (b) $\braket{\hat{Q}_{22}}$ along the LAP of $^{332}$Hs, calculated for different numbers of neutron and proton single-particle wave functions. The black line corresponds to $n_{\rm wf}^\pi=380$ and $n_{\rm wf}^\nu=620$, representing the well-converged reference case.}
\end{figure}

The primary source of the discrepancy in the collective mass lies in the values of the components of the inertia tensor. To illustrate it, in Fig.~\ref{M_tensor}, we show the diagonal components of the inertia tensor for two different numbers of proton single-particle wave functions, $n_{\mathrm{wf}}^{\pi}=300$ and $340$ (blue and red, respectively), and three values of $n_{\mathrm{wf}}^{\nu}$ (520, 560, 600) indicated by different markers. Changes in the tensor components are more pronounced when increasing $n_{\mathrm{wf}}^{\nu}$, indicating that $n_{\mathrm{wf}}^{\pi}=300$ is sufficient to obtain converged proton contributions. The reduction becomes negligible when 600 neutron wave functions are included. Hence, we established a heuristic relationship between the number of particles and the number of wave functions required to achieve well-converged MOI without incurring excessive computational cost:

\begin{align}
n_{\rm wf}^{\nu} &=
\begin{cases}
440, & N < 170 \\[6pt]
520, & 170 \le N < 210 \\[6pt]
N + 380, & N \ge 210
\end{cases}
\\[10pt]
n_{\rm wf}^{\pi}&=
\begin{cases}
260, & Z < 108 \\[6pt]
300, & 108 \le Z \le 114 \\[6pt]
Z+ 220, & Z > 114
\end{cases}
\end{align}

\subsection{Single versus double path calculation} \label{single_double}

\begin{figure}[htbp]
\begin{center}
\includegraphics[width=\linewidth]{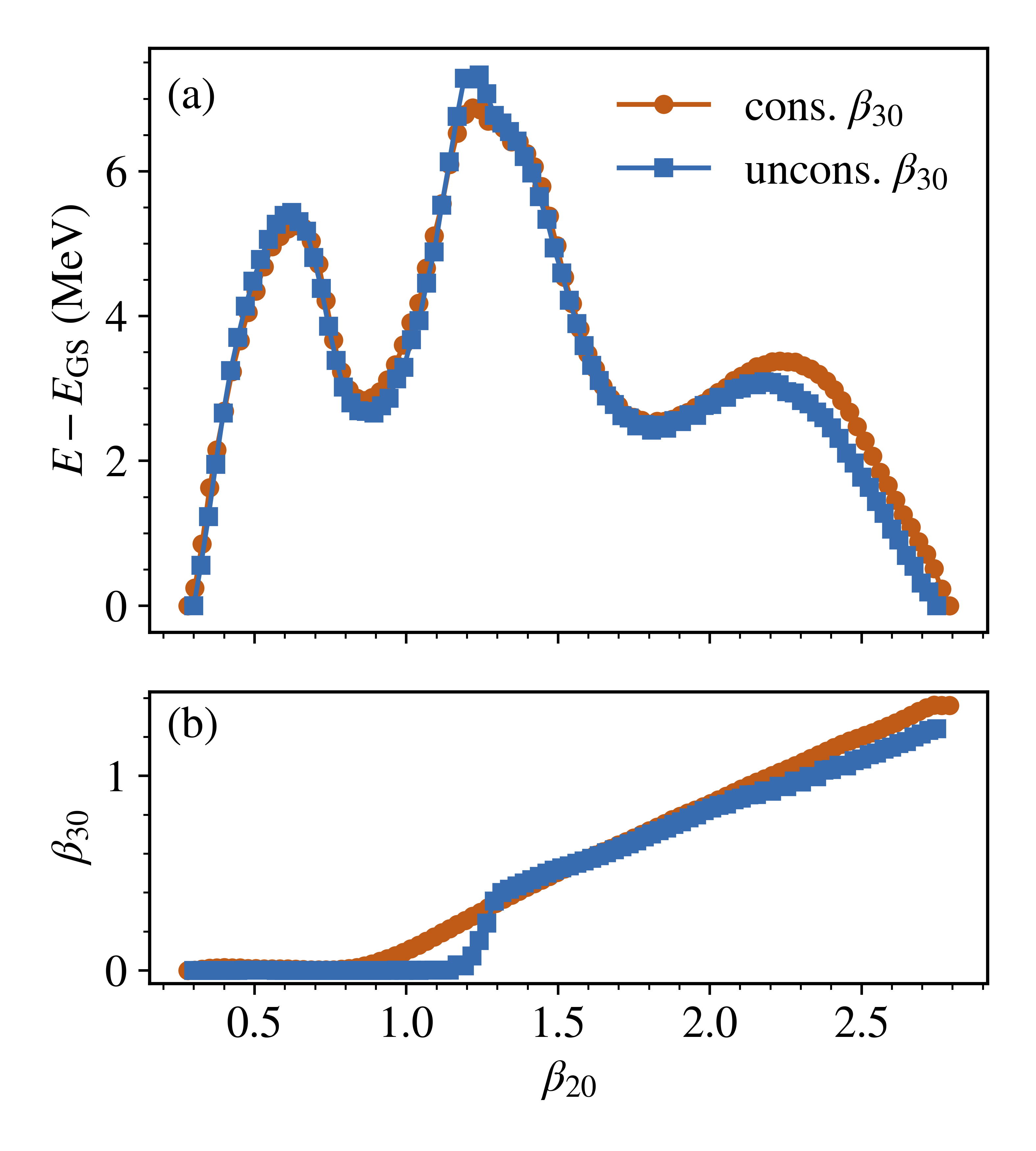}
\end{center}
\caption{\label{U244_paths} (a) Least-action fission paths of $^{244}$U across the $(\beta_{20}, \beta_{22})$ PES with and without constraining $\beta_{30}$ and (b) the evolution of $\beta_{30}$ in function of $\beta_{20}$.}
\end{figure}

To evaluate the impact of the double path methodology on the fission path topography, we present in Fig.~\ref{U244_paths}(a), the energy evolution along the two different LAPs corresponding to the PESs shown in Fig.~\ref{U244_CCU_CCC}, i.e., with and without the octupole constraint. We find that constraining $\beta_{30}$ has little effect around the first barrier. However, in the region of the second barrier—where the discontinuity was observed—the sharp peak disappears, leading to a reduction in the barrier height of approximately 450 keV. Conversely, the constraint causes the third hump to be about 200 keV higher than in the unconstrained case. Panel (b) of Fig.~\ref{U244_paths} shows the evolution of the octupole deformation as a function of $\beta_{20}$. The constrained path exhibits a smoother profile, consistent with the MEP on the $(\beta_{20}, \beta_{30})$ surface, while the unconstrained case shows a sharp increase around $\beta_{20} \approx 1.25$. In regions where the octupole deformation is similar in both cases—such as near the second and third barriers—the energy profiles also coincide. In terms of the action integral, we found a difference of approximately 1$\hbar$ between the two cases analyzed, with the $\beta_{30}$-constrained path exhibiting the larger action.

Another important aspect concerns the prescription used for the octupole deformation, namely whether it should be constrained to the values along the MEP or the LAP. On the one hand, adopting the MEP appears natural, as these values correspond to the unconstrained minima of the PES arising from the variational character of the HFB method. On the other hand, the physically relevant trajectory for fission dynamics is the LAP traced on the triaxial surface, which suggests that prescribing the octupole evolution based on the LAP may be more appropriate.

\begin{figure}[htbp]
\begin{center}
\includegraphics[width=\linewidth]{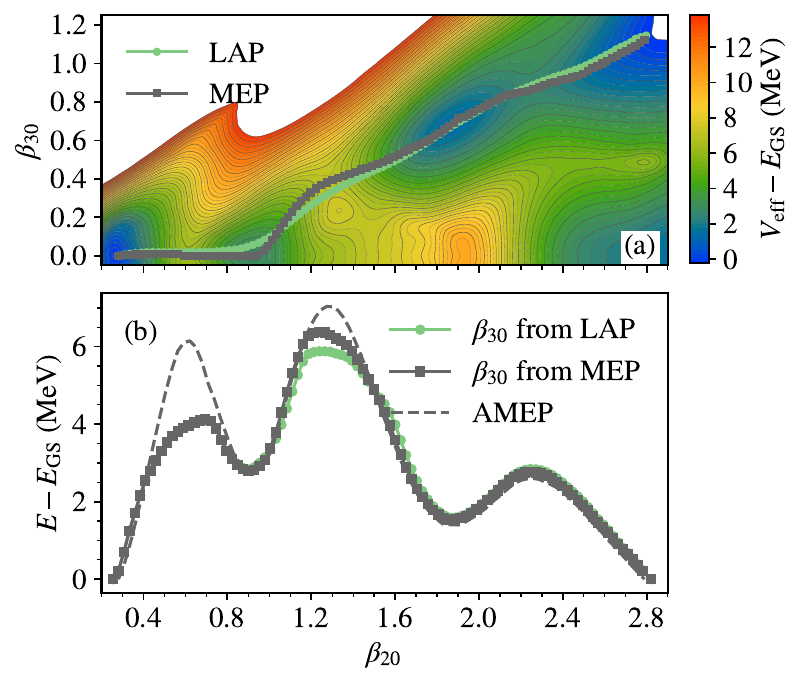}
\end{center}
\caption{\label{U250_lap_mep} (a) LAP and MEP across the $(\beta_{20},\beta_{30})$ plane. (b) Comparison of the $^{250}$U LAP in the $(\beta_{20},\beta_{22})$ plane obtained by constraining $\beta_{30}$ to follow either the MEP or the LAP from the previous surface. The axial minimum energy path (AMEP), is also included.}
\end{figure}

In Fig.~\ref{U250_lap_mep}, we illustrate the impact of these two choices for $^{250}$U. Panel (a) shows the axial quadrupole--octupole PES, together with both the MEP and the LAP extracted from it. Panel (b) displays the resulting triaxial LAP when the octupole deformation is constrained to follow the corresponding axial trajectories. For reference, we also include the conventional energy profile along the axial MEP. Somewhat counterintuitively, constraining the octupole deformation along the $(\beta_{20},\beta_{30})$ MEP leads to a higher first fission barrier, whereas the LAP-based prescription lowers it by up to 600~keV. In general, it is difficult to anticipate which prescription will produce the lowest barrier, since the comparison relies on the octupole evolution along an axial path, while the optimal triaxial trajectory may correspond to a different evolution of $\beta_{30}$. Although a fully self-consistent three-dimensional PES would ultimately remove this ambiguity, we find that in this case the difference in the action remains modest: 54.75\,$\hbar$ for the LAP-constrained trajectory compared to 54.06\,$\hbar$ for the MEP-constrained one.

\subsection{PyNEB sensitivity tests} \label{pyneb_test_section}

The nudged elastic band (NEB) method determines the fission trajectory by representing the path between the inner and outer turning points as a discrete chain of points (images, in its original terminology) connected by artificial spring forces. During the minimization, the physical force derived from the PES acts perpendicular to the path, driving the images toward the minimum-energy or least-action trajectory, while the spring force acts tangentially to maintain a regular distribution of points along the path~\cite{Flynn2022}. 

The spring-force constant $k$ therefore controls the balance between the smoothness of the path and the ability of the images to relax toward the optimal trajectory. If $k$ is too small, neighboring images drift unevenly and cluster in selected regions of the PES, leading to unstable or poorly resolved paths. In contrast, excessively large values of $k$ tend to overconstrain the chain, reducing the flexibility of the path and potentially generating numerical instabilities. The numerical behavior of the PyNEB also depends on the resolution of the underlying PES. Since the PES and inertia are known only on a finite deformation grid, increasing the number of images beyond the grid resolution introduces interpolation noise and enhances the sensitivity of the minimization procedure to local numerical fluctuations.

Owing to the exponential dependence of the penetration probability on the action, even small variations in the value of the latter can lead to significant deviations in the predicted SF half-lives. To ensure that the minimization yields the path with the smallest barriers (for MEPs) or the smallest actions (for LAPs), we perform up to 20 PyNEB calculations, depending on the complexity of the PES, using different exit points along the $ E = E_{\mathrm{GS}} $ energy contour and the $ \langle \hat{Q}_{\rm neck} \rangle = 6$ neck contour. To avoid artificially constraining the path, the initial point is fixed, while the outer turning point is allowed to vary.

\begin{figure}[htbp]
\begin{center}
\includegraphics[width=\linewidth]{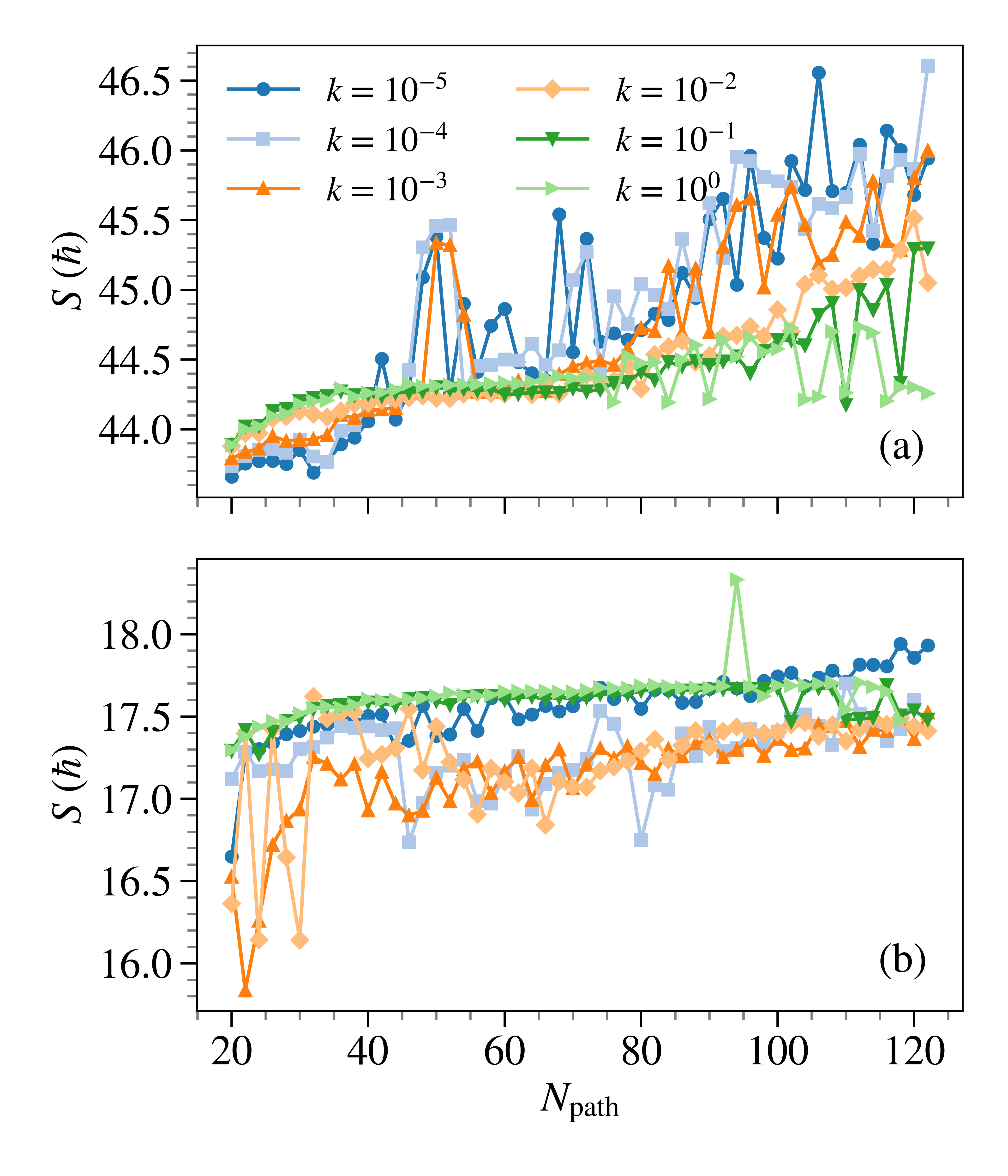}
\end{center}
\caption{\label{pyneb_test} Action evaluated along the path found by PyNEB of (a) $^{240}$Pu and (b) $^{286}$Fl as a function of the number of discretization points along the path for different values of the spring-force constant $k$.}
\end{figure}

When this strategy is applied in the $(\beta_{20}, \beta_{30})$ plane, it determines the fission mode of the system, namely whether the process is mass-symmetric or mass-asymmetric, which are typically associated with different barrier heights and fission times. Although these differences are physically justified, for a given path with fixed $s_{\rm in}$ and $s_{\rm out}$, different internal PyNEB parameters may still yield different values of the action. To quantify this effect, we analyze here the impact of the spring-force strength $k$ and the number of discretization points along the fission path, $N_{\rm path}$. Specifically, we computed the action along the LAP of $^{240}$Pu and $^{286}$Fl in the $(\beta_{20}, \beta_{30})$ collective space for different values of $k$ and $N_{\rm path}$. The corresponding results are shown in Fig.~\ref{pyneb_test}(a) and (b). In all cases, the inner and outer turning points were fixed to the same values, namely $s_{\rm in} = (0.28, 0.00)$ and $s_{\rm out} = (2.16, 0.96)$ for $^{240}$Pu and $s_{\rm in} = (0.17, 0.00)$ and $s_{\rm out} = (0.86, 0.01)$ for $^{286}$Fl.

In the case of $^{240}$Pu, we observe that for the smallest values of $k$ the action can vary by more than two units, showing a strong sensitivity to the number of points used to discretize the path. Increasing the spring strength generally leads to more stable results; however, excessively large values of $k$ can introduce numerical instabilities when using a large number of points. In such cases, convergence typically requires a substantially larger number of iterations and may fail altogether. For intermediate spring constants, $10^{-2} \leq k \leq 1$, we identify a clear plateau for paths containing between 30 and 90 points, over which the action varies by less than one unit. If a larger number of points is used (more than 120), numerical noise is introduced due to the grid resolution of the PESs and the path calculation usually did not converge.

In the case of $^{286}$Fl, we found two distinct types of fission pathways emerging. For $k = 10^{-5}, 10^{-1},$ and $1$, most values of $N_{\rm path}$ converged to a mass-symmetric outer turning point, located very close to the initial $s_{\rm out}$. In contrast, for $k = 10^{-4}, 10^{-3},$ and $10^{-2}$, the calculations converged to a mass-asymmetric outer turning point, yielding a path that closely follows the minimum-energy path obtained on the same potential-energy surface. Despite these qualitatively different endpoints, the corresponding actions differ by no more than $0.5\,\hbar$.This behavior illustrates the coexistence of multiple stationary-action paths in the multidimensional collective space as mentioned in Section~\ref{th_fis_paths}. The fact that qualitatively different paths can yield similar actions also underscores the limitations inherent in reducing the multidimensional fission dynamics to a single one-dimensional trajectory, as required for interfacing microscopic calculations with reaction codes~\cite{Sanchez2026c}.

In all cases considered, we found that $k = 1$ provides the most stable results; accordingly, we kept this value fixed for the large-scale calculations. Regarding the number of discretization points along the path, we employed 100 points for longer paths---i.e., those reaching $\beta_{20} > 2.5$, as in the actinide region. For shorter paths, such as those typical of superheavy nuclei, we used between 40 and 60 points.

\section{Benchmarking \texorpdfstring{\NoCaseChange{BSkG3}}{BSkG3}}
\label{section_IV}

\begin{figure}[htbp]
\begin{center}
\includegraphics[width=\linewidth]{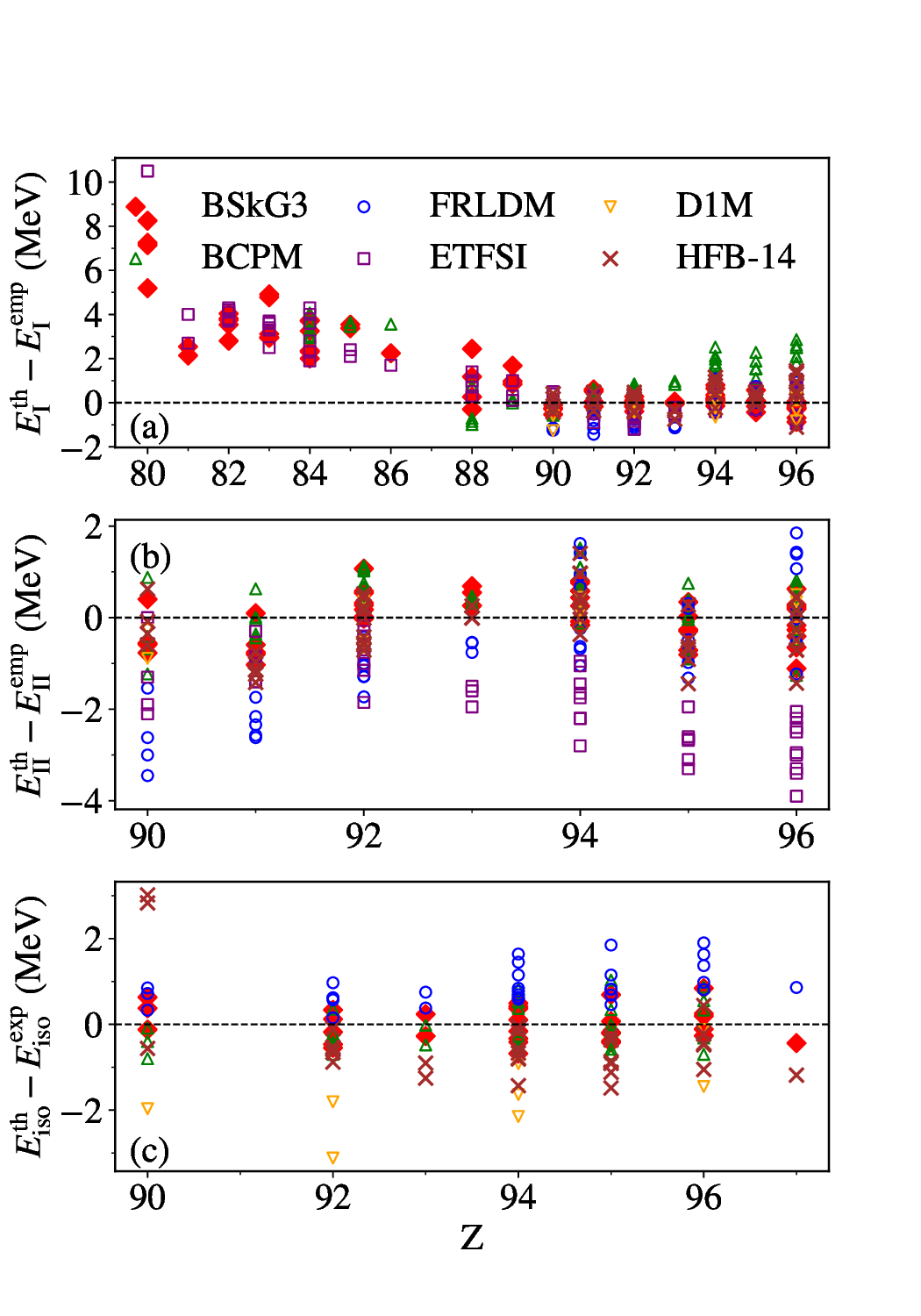}
\end{center}
\caption{\label{EI_EII_Eiso_comp} Deviations of the (a) primary, (b) secondary fission barriers and (c) isomer excitation energies obtained with different models w.r.t. the RIPL-3 empirical values. For nuclei with $Z$\textless 90, empirical data are available for only one barrier.}
\end{figure}

Before presenting the large-scale results, we first assess the performance of our model relative to existing approaches in reproducing key fission properties. We focus specifically on three quantities that are routinely used as benchmarks in theoretical studies: empirical fission-barrier heights, isomer excitation energies and SF half-lives. Figures~\ref{EI_EII_Eiso_comp}(a)-(c) display model-to-data comparisons for 77 primary and 45 secondary empirical barriers and 30 fission isomers, respectively, as compiled in the RIPL-3 database~\cite{capote2009}. Summary statistics for elements with $Z \geq 90$, including mean and rms deviations, are provided in Table~\ref{tab:barrier_deviations}.  For context, we include results from three microscopic frameworks: The Barcelona-Catania-Paris-Madrid (BCPM) functional~\cite{Baldo13,Giuliani18b}, the D1M Gogny functional~\cite{Goriely09,Lemaitre2018} (restricted to even-even nuclei), and HFB-14~\cite{Goriely07}---alongside predictions from the macroscopic-microscopic finite-range liquid-drop model (FRLDM)  \cite{moller2009,Moller15} and the extended Thomas-Fermi plus Strutinsky Integral (ETFSI)  \cite{Mamdouh98,mamdouh2001} model. For the fission isomers, we only consider the well between the first two barriers, regardless of the presence of additional barriers and isomers. The values predicted by BSkG3 are those obtained along the LAP in the $(\beta_{20}, \beta_{22})$ plane; hence, they differ slightly from those reported in \cite{Grams23}, which were obtained as local extrema on the PESs.

In Fig.~\ref{EI_EII_Eiso_comp}(a), a notable improvement in model accuracy for the fission barriers is observed for nuclei with $Z \geq 88$ compared to those with $Z < 88$, with deviations remaining below 3.5~MeV. In this region, the performance of the different models is relatively comparable. BCPM, however, exhibits the largest discrepancies, particularly for $Z \geq 94$, which is unsurprising given that it was not fitted to reproduce these empirical barriers and preserves axial symmetry without triaxial corrections, unlike D1M. In contrast, HFB-14 and BSkG3 incorporate collective corrections tuned to match to empirical barrier values. For pre-actinide nuclei, although they were excluded in the fitting protocol, BSkG3 performs comparably to ETFSI and BCPM. The systematic large deviations observed in this region--where the predicted barriers are also significantly higher--across these models indicate a significant shortcoming in the description of fission in light nuclei—an issue that is unlikely to be resolved merely by improving the parameter adjustment of the models. Fortunately, due to the low fission probability of such nuclei the impact of our overestimation in the r-process is expected to be low.

Regarding the secondary barrier, most models exhibit larger deviations than for the primary barrier, with the exception of BCPM and D1M. Overall, microscopic models provide more accurate secondary barriers than those obtained with macroscopic--microscopic approaches. Notably, the ETFSI set shows an rms deviation of 2.03~MeV---around four times larger than that of BSkG3---and its predicted barriers systematically decrease relative to the empirical values as $Z$ increases. Only the D1M secondary barriers achieve a lower rms value, 0.43~MeV, although this result is based on a data set that is approximately three times smaller. 

For the fission isomer excitation energies, the deviations predicted by most models are generally smaller than those observed for the barrier heights. The largest discrepancies occur for $^{230,231}$Th in the HFB-14 predictions and for $^{234}$U in D1M, with  differences exceeding 2 MeV. BPCM is capable to reproduce these values with an rms deviation as small as 0.52 MeV, while D1M reported its worst performance, with an rms deviation of 1.84 MeV. Again, among all these models, BSkG3 results in the smallest deviations, with a mean deviation of less than 40 keV and an rms of 0.38 MeV.

\begin{table}[h!]
\centering
\caption{Mean (theory-empirical) and rms deviations of theoretical fission barriers and isomers w.r.t. the 45 RIPL-3 empirical values for $Z \geq 90$. ETFSI predictions are not available for the isomers.}
\label{tab:barrier_deviations}
\begin{tabular}{lccccccc}
\hline
\hline
\multirow{2}{*}{Model} & \multicolumn{2}{c}{$E_{\rm I}$ (MeV)} & \multicolumn{2}{c}{ $E_{\rm II}$ (MeV)} &  \multicolumn{2}{c}{ $E_{\rm iso}$ (MeV)}  \\
 & mean & rms & mean & rms &  mean & rms \\
\hline
BCPM     &  1.06 & 1.41  &  0.30 & 0.72 &  0.09 & 0.52 \\
BSkG3    & -0.01 & 0.32  &  0.00 & 0.54 &  0.04 & 0.38 \\
D1M      & -0.23 & 0.53  & -0.06 & 0.43 & -1.65 & 1.84 \\
ETFSI    & -0.04 & 0.57  & -1.77 & 2.03 &       &      \\
FRLDM    & -0.22 & 0.81  & -0.66 & 1.41 &   0.9 & 1.00 \\
HFB-14   &  0.27 & 0.60  & -0.20 & 0.69 & -0.40 & 1.07 \\
\hline
\hline
\end{tabular}
\end{table}

\begin{figure}[htbp]
\begin{center}
\includegraphics[width=\linewidth]{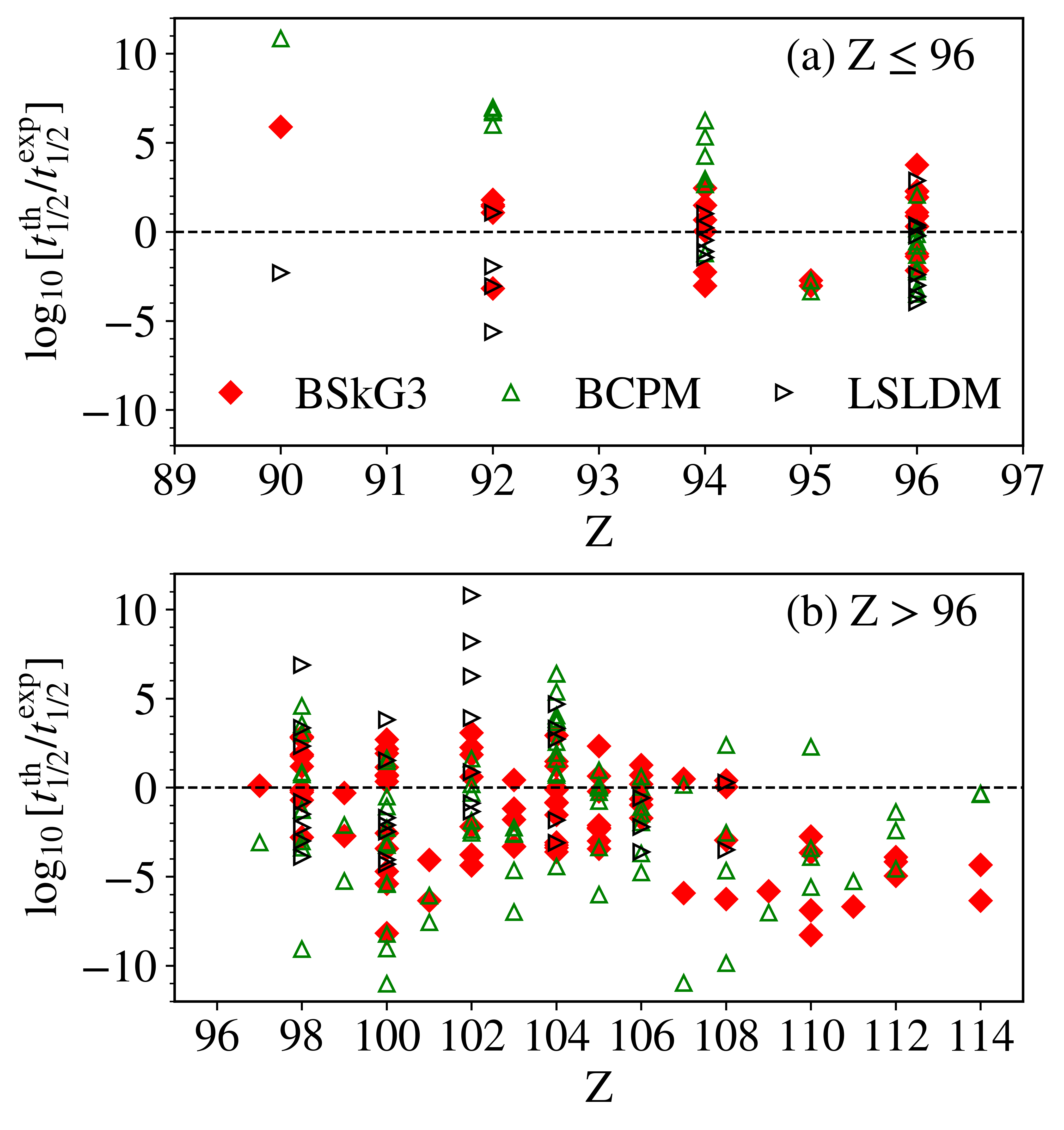}
\end{center}
\caption{\label{spf_diff_models} Deviations of the GS SF half-lives obtained with different models w.r.t. the experimental values.}
\end{figure}

We now examine the accuracy of various models in predicting GS SF half-lives. Although our previous work primarily focused on this analysis \cite{Sanchez2026a}, we highlight a few insights. In line with Fig. \ref{EI_EII_Eiso_comp}, Fig. \ref{spf_diff_models} presents the deviations of the different models with respect to the experimental values reported in Ref.~\cite{Kondev21}. Given that these deviations can span roughly ten orders of magnitude, the results are displayed on a logarithmic scale. Providing a comprehensive theoretical description of SF remains extremely challenging, and only a few models have produced results of comparable scope. For this reason, we only include here the theoretical results that can be meaningfully compared with ours. In particular, we consider the BCPM functional and the Lublin–Strasbourg liquid-drop model (LSLDM) \cite{Pomorski03,Blanco23}. These models represent the least phenomenological approaches available in the literature. Another class of results can be found from empirical formulas for the half-lives, which depend on the height and curvature of the fission barriers. However, these formulas are explicitly fitted to reproduce the experimental half-lives and their performance depends on the specific model used to compute the barriers and on the assumed presence or absence of an outer barrier. For instance, Ref.~\cite{Khuyag20} uses two different expressions, depending on whether the barriers were taken with FRLDM or ETFSI. Moreover, since these approaches are tailored specifically to SF half-life predictions, they cannot be straightforwardly extended to describe other fission observables. BSkG3 predictions are found to differ from such empirical formulas by many orders of magnitude for experimentally unknown half-lives, as discussed in Sect.~\ref{sec:compother}.

For a clearer visualization, Fig. \ref{spf_diff_models}(a) displays the results for nuclei with $90 \le Z \leq 96$—the region encompassing the fission barriers considered in the BSkG3 fitting protocol—while panel (b) shows the results for nuclei with $Z > 96$. In general, BSkG3 provides the closest overall agreement with experiment, followed by LSLDM (both within three orders of magnitude) and BCPM (four orders of magnitude). For ease of comparison, Table~\ref{tab:sf_deviations} lists the mean and rms deviations, $f_{\rm mean}$ and $f_{\rm rms}$, computed as in Ref.~\cite{Sanchez2026a}. It is important to note the substantial differences in size of the datasets: LSLDM contains only 50 values, whereas BSkG3 and BCPM include 107 SF half-lives.

In Fig.~\ref{spf_diff_models}(a), all three models exhibit broadly comparable performance, with similar deviations across the dataset; the largest discrepancy occurs for $^{232}$Th. Figure~\ref{spf_diff_models}(b) displays a slightly wider dispersion than observed in panel (a), together with a systematic underestimation for the heavier nuclei, starting at $Z\sim 107$. In particular, both BCPM and BSkG3 yield a mean deviation $f_{\rm mean}\sim 10^{-2}$ for the subset with $Z>96$. BSkG3 once again yields the smallest rms deviations—both globally and within each subset—despite the fact that its parameters were not further adjusted \textit{a posteriori} to reproduce SF half-lives. The BSkG3 fit was performed only once, using masses, radii, and fission-barrier heights as constraints, with no subsequent tuning. This stands in contrast to BCPM and LSLDM, for which the effective inertias were modified to better reproduce the experimental half-lives.

\begin{table}[htbp]
\centering
\caption{Deviations of theoretical SF half-lives relative to experiment. Note that deviations for LSLDM concern only 50 SF half-lives, whereas those for BSkG3 and BCPM include all the 107 SF half-lives experimentally available.}
\setlength{\tabcolsep}{0.5pt}
\begin{tabular}{lcccccc}
\hline
\hline
Model & \multicolumn{2}{c}{All Z} & \multicolumn{2}{c}{Z $\le$ 96} & \multicolumn{2}{c}{Z $>$ 96} \\
 & $f_{\rm mean}$ & $f_{\rm rms}$ & $f_{\rm mean}$ & $f_{\rm rms}$ & $f_{\rm mean}$ & $f_{\rm rms}$ \\
\hline
BCPM    & $1.2\!\times\!10^{-1}$ & $1.9\!\times\!10^{4}$ & $8.9\!\times\!10^{1}$ & $3.8\!\times\!10^{4}$ & $1.8\!\times\!10^{-2}$ & $1.5\!\times\!10^{4}$ \\
BSkG3   & $7.6\!\times\!10^{-2}$ & $1.1\!\times\!10^{3}$ & $2.6\!\times\!10^{0}$ & $2.2\!\times\!10^{2}$ & $2.8\!\times\!10^{-2}$ & $1.7\!\times\!10^{3}$ \\
LSLDM   & $7.1\!\times\!10^{-1}$ & $2.8\!\times\!10^{3}$ & $5.1\!\times\!10^{-2}$ & $2.8\!\times\!10^{2}$ & $3.1\!\times\!10^{0}$ & $7.9\!\times\!10^{3}$ \\
\hline
\hline
\end{tabular}
\label{tab:sf_deviations}
\end{table}

\begin{figure*}[htbp]
\begin{center}
\includegraphics[width=\linewidth]{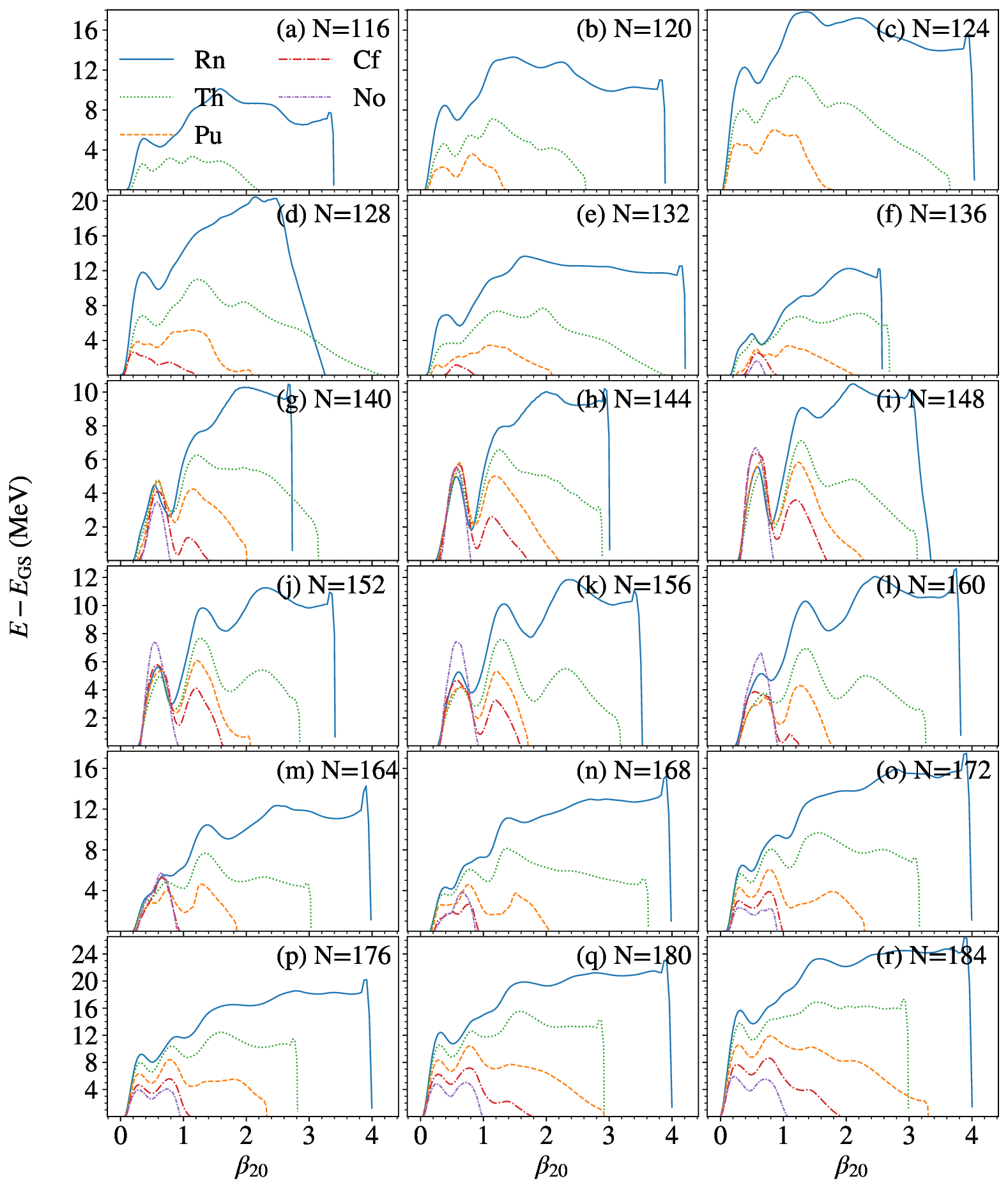}
\end{center}
\caption{\label{fission_paths_N} Selected collection of one-dimensional fission paths in $\beta_{20}$ for different isotopes of Rn, Th, Pu, Cf and No for $N\in[116,184]$ obtained with BSkG3.}
\end{figure*}

It is worth noting that BCPM and BSkG3 employ microscopic collective inertias derived self-consistently from their underlying energy-density functionals, whereas LSLDM relies on the irrotational flow model. Any modification to the inertia scheme has the potential to alter the predicted half-lives, and addressing these uncertainties will require dedicated future work. For the time being, however, the validation of our approach must rely on comparisons with available observables. The level of agreement achieved in barrier heights, isomer excitation energies and SF half-lives, indicates a consistent and sufficiently accurate description to justify extrapolations into more exotic regions of the nuclear chart.

\section{Global overview of fission paths}\label{section_V}

\subsection{General properties: number of humps and dependency on
\texorpdfstring{$N$ and $Z$}{N and Z}}

We now examine the topography of the fission paths predicted by the BSkG3 model in the region $Z \in [80,118]$ extending from the predicted proton to the neutron drip lines.

Figure~\ref{fission_paths_N} illustrates the evolution of one-dimensional fission paths along $\beta_{20}$ for five selected isotopic chains—Rn, Th, Pu, Cf, and No—across different neutron numbers. For all displayed neutron numbers, the Rn isotopes do not exhibit a semiclassical outer turning point. Instead, the fission paths increase almost monotonically in energy until scission due to the strong shell effects of the pre-fragments \cite{Bernard24}. Similar topographies are observed for the lightest actinides, see Fig.~\ref{fission_paths_N}(m)–(r),particularly as one approaches the neutron shell closure at $N=184$. Because the energy at scission decreases with increasing $Z$, nuclei heavier than Am in the vicinity of the shell closure reach scission below the ground state energy. This allowed us to find an exit point at $E = E_{\rm GS}$ for the evaluation of the tunneling probability.

In Fig.~\ref{fission_paths_N}, BSkG3 predicts triple-humped fission paths for the Th isotopic chain. While experimental evidence generally indicates this feature around $N \sim 140$ \cite{Gavron76,Blons84}, BSkG3 shows a rather flat plateau after the second barrier, followed by a gradual decrease in energy, in agreement with predictions obtained using SkM* \cite{McDonnell13}. As the neutron number increases, the depth of the second well becomes more pronounced, giving rise to a third barrier whose shape gradually resembles that of the inner and outer barriers--see Fig.~\ref{fission_paths_N}(i)–(l).

\begin{figure}[htbp]
\begin{center}
\includegraphics[width=1.0\linewidth]{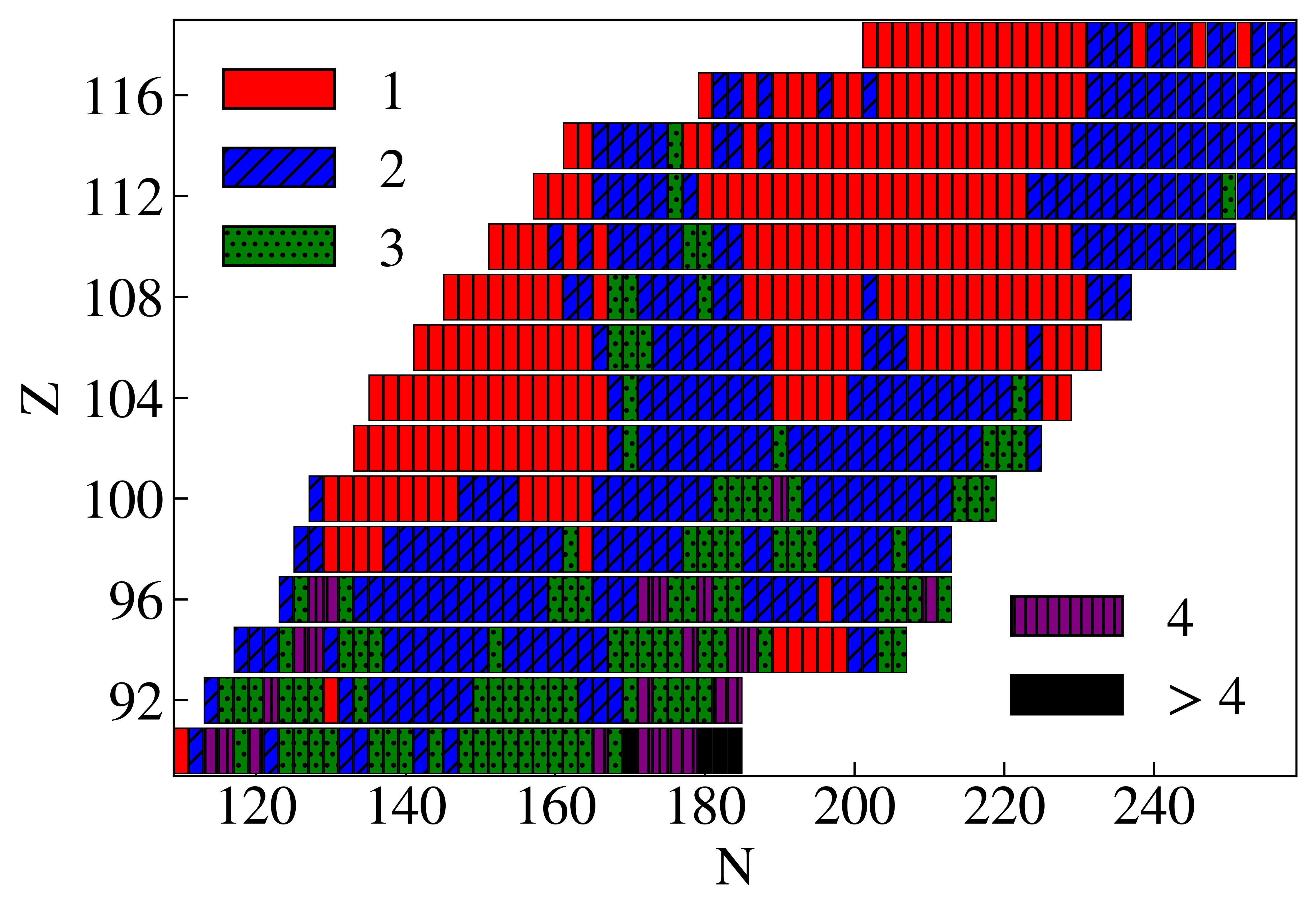}
\end{center}
\caption{\label{barrier_count} Number of fission barriers along the LAPs for the even-even nuclei in the region  $90 \leq Z \leq 118$ obtained with BSkG3.}
\end{figure}

Increasing the number of protons does not significantly affect the position or shape of the barriers up to the second barrier; however, the subsequent energy decrease becomes more pronounced due to stronger Coulomb repulsion. Consequently, the number of humps that can be found above the GS decreases with increasing $Z$ for a given $N$. To illustrate this, Fig.~\ref{barrier_count} shows the number of humps for even-even nuclei with $Z \geq 90$. The saddles were identified by locating local maxima, analyzing the energy variation over five points to the left and right of each point along the fission path, separated between 0.01 and 0.03 in $\beta_{20}$. Although some barriers may be missed due to the flatness of the curves, or irregularities in the PESs, Fig.~\ref{barrier_count} captures the general trend in the number of barriers. The conventional double-humped structure appears in roughly half of the nuclei; including not only the usual actinide region with $N\sim140$, but also heaviest elements of the nuclear chart ($Z>108,\; N>220$). The rest of the region exhibit either single-humped paths--mainly in the neutron-deficient region for $Z > 100$ and $N \in [180,230]$--or more exotic cases with three, four, or more barriers, particularly among the lighter nuclei.

\subsection{Deformation at the saddles: the role of triaxiality and mass-assymetry}

\begin{figure*}[htbp]
\begin{center}
\includegraphics[width=\linewidth]{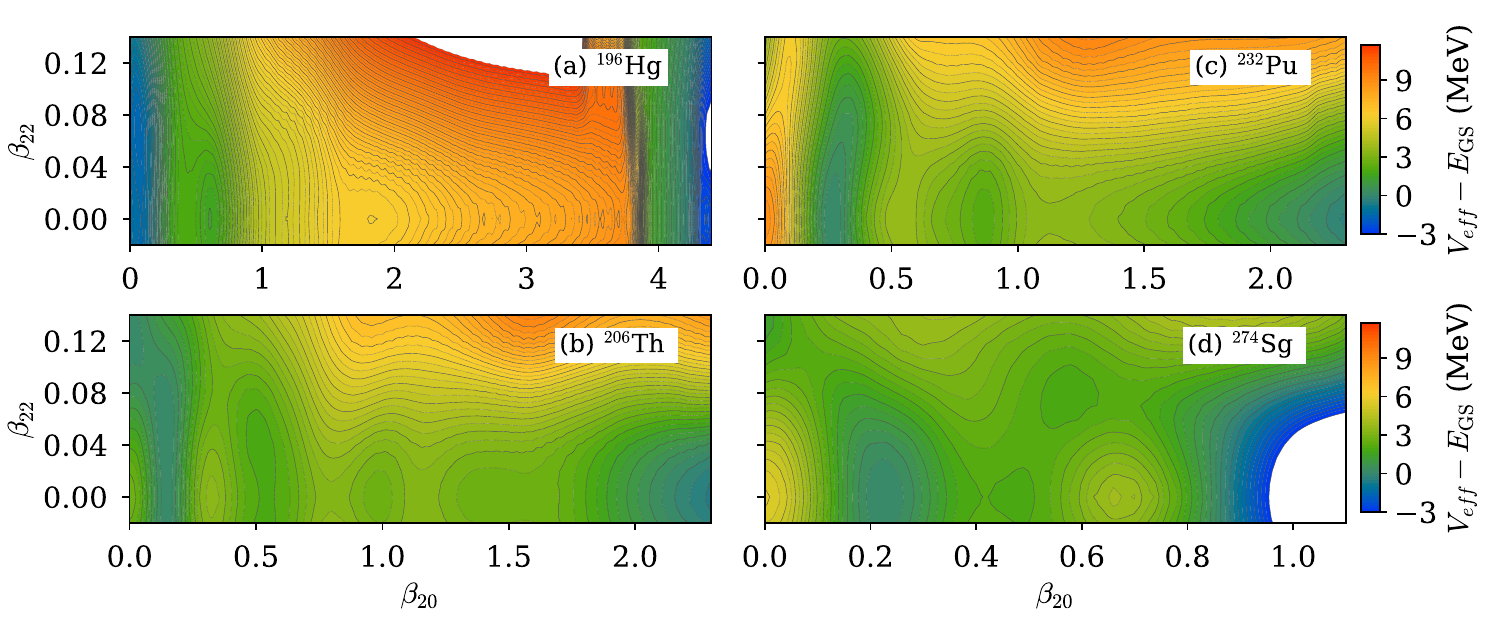}
\end{center}
\caption{\label{4PESs} PES in $(\beta_{20},\beta_{22})$ for four different nuclei obtained with BSkG3. The color scale for $^{196}$Hg has been extended to cover energies up to 20 MeV.}
\end{figure*}

Based the theoretical studies of actinides \cite{Schunck14,Schunck16,Ryssens23,Erler12,Rodriguez14}, it is generally accepted that the inner fission barrier is affected by triaxial deformation, while the outer barrier by octupole deformation. However, this behavior does not necessarily hold as a universal rule. 

Figure~\ref{4PESs} presents four representative PESs that illustrate distinct topological features in terms of barrier multiplicity and triaxiality. Specifically, both $^{206}$Th and $^{232}$Pu exhibit the typical actinide behavior. In contrast, $^{196}$Hg shows no triaxial saddles, while $^{274}$Sg presents a pronounced triaxial deformation near the second barrier, around $\beta_{20} \approx 0.7$, whereas the first barrier at $\beta_{20} \approx 0.4$ remains essentially axial.

\begin{figure*}[htbp]
\begin{center}
\includegraphics[width=\linewidth]{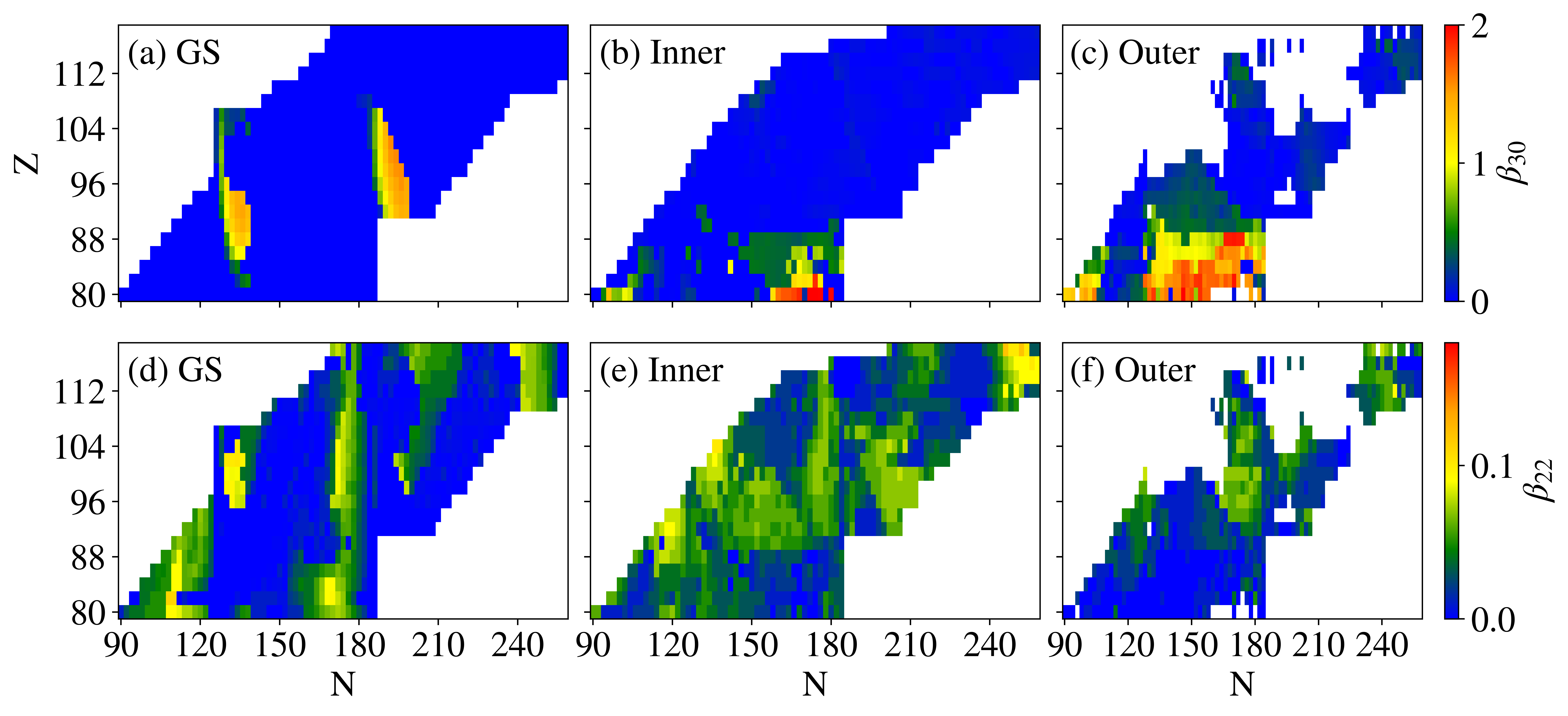}
\end{center}
\caption{\label{def_barr} Octupole deformation $\beta_{30}$ for the (a) GS, (b) inner barrier, and (c) outer barrier, and triaxial deformation $\beta_{22}$ for the (d) GS, (e) inner barrier, and (f) outer barrier. The outer barrier corresponds to the highest saddle point located beyond the first isomeric potential well. Because the GS exhibits substantially smaller octupole deformation, the color scale in panel (a) was reduced by a factor of 10 relative to panels (b) and (c).}
\end{figure*}

In Fig.~\ref{def_barr}, we present these collective degrees of freedom for the GS, inner and outer barriers for the even-even cases. Given the possible multiplicity of outer barriers, we define the outer barrier as the highest saddle point located after the first local minimum, ensuring a consistent comparison throughout the region. 

Actinides with $N \in [130,150]$ exhibit the expected behavior: low or negligible octupole deformation ($\beta_{30}$) and $\beta_{22}\approx 0.05$ at the inner barrier, together with a mild octupole deformation ($\beta_{30} \approx 0.30\text{–}0.40$) and an almost axially symmetric shape at the outer barrier. However, this characterization does not hold uniformly across the entire region we explored. In Fig.~\ref{def_barr}(b), the inner barrier shows negligible octupole deformation for $Z \geq 90$, whereas in the neutron-rich pre-actinide region both inner and outer barriers become strongly octupole deformed, reaching values as large as $\beta_{30} = 2.0$. Because their paths increase in energy almost monotonically, most inner barriers occur at relatively large deformations, which complicates direct comparisons. This is illustrated in Fig.~\ref{fission_paths_N}(n), where the inner barrier of $^{254}$Rn corresponds to the hump near $\beta_{20} \sim 1.5$, although two smaller, flatter humps can also be identified at $\beta_{20} \sim 0.4$ and 0.8. We observe no correlation between the octupole deformation of the GS and the fission barrier properties. In nuclei with octupole-deformed GS configurations, shown in In Fig.~\ref{def_barr}(a), the MEP rapidly evolved toward mass-symmetric shapes at small quadrupole deformations, and $\beta_{30}$ increases only when such configurations become energetically favorable, which occurs primarily at the outer barrier.

In the superheavy region ($Z>108,\; N>220$), the second saddle is nearly mass-symmetric but retains a non-negligible triaxial deformation, in agreement with predictions earlier studies~\cite{cwiok1996a,kowal2010,Abusara12,Lu14,Afanasjev18}.

In Fig.~\ref{def_barr}(e), we observe that noticeable $\beta_{22}$ values are present throughout the region, with the exception of the heaviest pre-actinide isotopes and three areas centered around (i)$Z \approx 104, N \approx 160$; (ii) $Z \approx 112, N \approx 190$; and  (iii) $Z \approx 110, N \approx 230$. We also identify a band within $N\in[170, 190]$ that narrows with increasing $Z$, maintaining an almost constant triaxial deformation. Furthermore, the superheavy region ($Z > 110$, $N > 230$) exhibits the most triaxially deformed inner barriers, with $\beta_{22}$ reaching $0.1$. By contrast, the outer barriers along the LAPs show non-negligible triaxility in three major regions: one near $N=126$ in the proton-rich actinide region, another spanning $N\in[170,190]$, starting at $Z=94$, and a smaller region at $N\in[230,250]$ for elements with $Z>108$. Moreover, these regions coincide with those in which the GSs are triaxially deformed, as shown in Fig.~\ref{def_barr}(d) for nuclei beyond Th. However, in the pre-actinide region, nuclei with triaxially deformed GSs evolve toward axially symmetric inner barriers, whereas axially symmetric GSs generally give rise to triaxial inner barriers.

The impact of triaxiality on the energies of the inner and outer fission barriers has been discussed in previous studies, notably Ref. \cite{Ryssens23}, which analyzed 45 nuclei from the RIPL-3 dataset. Consistent with \cite{Ryssens23}, we find that, locally—in the actinide region with experimentally known nuclei—triaxiality can reduce the barrier heights by up to a few MeV, although this effect does not hold globally. Moreover, a direct comparison between axial and triaxial fission paths is often non-trivial, as several distinct scenarios complicate the interpretation. These include: (i) the emergence of a local minimum at the top of a barrier, effectively splitting a single hump into two; (ii) the complete disappearance of a barrier due to its substantial lowering; (iii) the appearance of an additional barrier when the GS is triaxially deformed, and (iv) the presence of spurious barriers arising from numerical irregularities in the PESs. In light of these complexities, a systematic and detailed analysis of the approximately 3300 triaxial fission paths is not feasible.

\begin{figure}[htbp]
\begin{center}
\includegraphics[width=\linewidth]{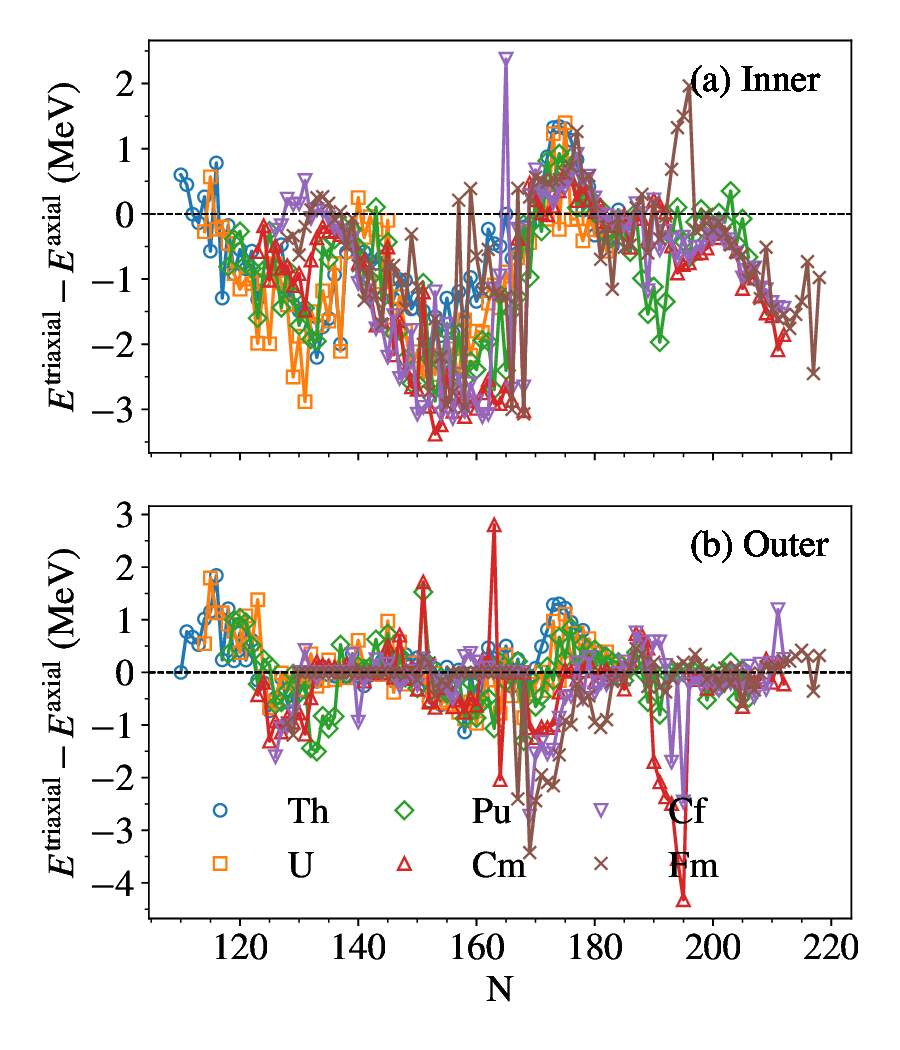}
\end{center}
\caption{\label{triax_chains} Energy difference between the triaxial and axial barriers for even-$Z$ isotopic chains between Th and Fm. The triaxial barriers were found along the LAP calculated on the $(\beta_{20},\beta_{22})$ plane, while the axial were those obtained along the axial line with $\beta_{22}=0$.
}
\end{figure}

Nevertheless, to illustrate the variety of scenarios, Fig.~\ref{triax_chains} summarizes the energy differences between the triaxial LAP and the axial path for the (a) inner and (b) outer barriers along the even-$Z$ isotopic chains from Th to Fm. In general, triaxiality has a stronger impact on the inner fission barrier, although this trend is not universal. In Fig.~\ref{triax_chains}(a), we observe that--largely independent of proton number--the region around $N \approx 150$ is most strongly affected, with energy reductions of up to 3.5 MeV, followed by $N \approx 130$, where reductions reach up to 3 MeV for the lightest actinides due to shell effects. In some cases, the barrier height even increases when the GS is triaxially deformed, particularly near $N = 174$. For the outer barrier, Fig.~\ref{triax_chains}(b) shows that the overall effect of triaxiality is weaker. However, in certain cases, the energy reduction reaches up to 4 MeV, as seen for Cm isotopes near $N \approx 190$. This is due to the emergence of a small additional barrier around $\beta_{20} \approx 0.2$ when triaxiality is included; in such cases, the analysis is misleading as different barriers at different $\beta_{20}$ are being compared.

\subsection{Primary fission barriers}

\subsubsection{Comparison between models}

We present the primary fission barriers ($B_f$) calculated along the LAPs using BSkG3 for nuclei in the region $80 \leq Z \leq 118$. By definition, the MEPs yield the lowest barriers on the PESs; however, the differences with the LAP results are typically within a few hundred keV. Notably, these discrepancies arise predominantly in the triaxial outer barriers, as the collective inertia tends to favor paths that are nearly axially symmetric. Consequently, and given the excellent agreement with the empirical values, we focus our analysis on the predictions obtained by minimizing the action integral.

\begin{figure*}[htbp]
\begin{center}
\includegraphics[width=\linewidth]{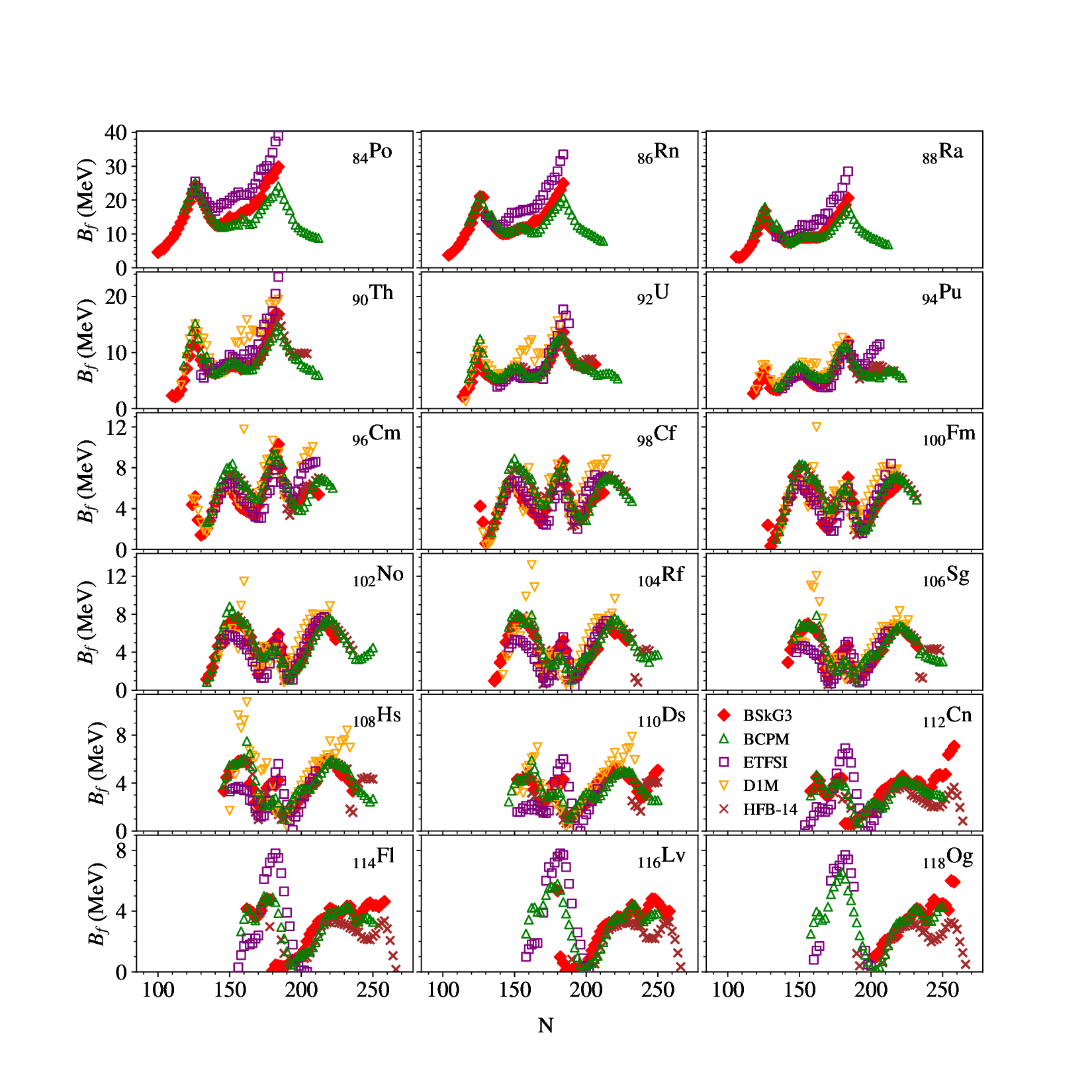}
\end{center}
\caption{\label{comp_barriers} Primary fission barrier heights, $B_f$, of nuclei in the region $80\leq Z \leq 118$ with $Z,N$ even obtained with BCPM, BSkG3, D1M, ETFSI and HFB-14.}
\end{figure*}

In Fig.~\ref{comp_barriers}, these barriers are compared with results obtained from BCPM, D1M, HFB-14, and ETFSI. For clarity, nuclei with odd $Z$ or $N$ have been omitted\footnote{Hg and Pb isotopes are excluded, as they could only be compared against ETFSI, with very limited data.}.
For the pre-actinide nuclei, BSkG3 reproduces the same trend as BCPM and ETFSI, exhibiting two pronounced peaks at the magic neutron numbers $N=126$ and $N=184$. For $N>140$, BSkG3 and BCPM remain relatively close, whereas ETFSI starts to deviate, reaching differences of more than 10~MeV compared to the other two models near the $N=184$ peak. However, the primary barriers in this region are rather sensitive to the choice of the ending point of the fission path, i.e., the value of the neck size that defines scission. Nonetheless, due to the high barriers of these nuclei, the results are not expected to significantly alter the description of the $r$-process.

For nuclei with $100 < Z < 110$, ETFSI predicts smaller barriers for neutron-deficient isotopes, whereas D1M yields higher barriers compared to BSkG3, BCPM, and HFB-14, exceeding those models by roughly 2~MeV across most of the isotopic chains. For instance, for $^{274}$Sg (shown in Fig.~\ref{4PESs}(d)), BCPM and HFB-14 predict $B_f = 4.45$ and 4.22~MeV, respectively, while BSkG3 yields a significantly lower barrier of 2.25~MeV. This difference is mainly attributed to the preservation of axial symmetry in BCPM. In the case of HFB-14, the triaxial correction was applied only to the inner barrier, which in this nuclide corresponds to the secondary barrier.

Another notable difference is the increase in the fission barrier heights for superheavy neutron-rich nuclides in comparison with BCPM and HFB-14. This is visible in nuclei with $Z\geq 110$, particularly for nuclei with $N > 220$. However, whether such nuclei can be produced in NSMs depends on the stability of lighter chemical elements. Nonetheless, if the neutron-capture process extends into this region, BSkG3 predicts superheavy nuclei to exhibit longer fission decay times than previously modeled~\cite{Goriely15}.

\subsubsection{Impact on the nucleosynthesis \texorpdfstring{$r$-process}{r-process}}

Of particular interest is the prediction from BSkG3, which yields smaller primary barriers for nuclei with $Z>94$ in the region $N \in [150,170]$ compared to other models. These nuclei are especially important because they directly influence the production of $^{254}$Cf during the late stages of the $r$-process~\cite{zhu2018}. This isotope, if produced, is known to contribute significantly to the late-time decay heat in kilonovae resulting from NSMs~\cite{Wanajo2018}, due to its characteristic measured SF half-live. For instance, BCPM (HFB-14) predicts $B_f(^{254}\text{Bk}) = 7.08\ (6.85)$ MeV, whereas BSkG3 yields a considerably lower value of 5.03 MeV. This reduced barrier accelerates the fission of $^{254}$Bk, thereby suppressing the production of $^{254}$Cf, as observed in preliminary $r$-process simulations. The overall impact of this behavior in the actinide region will be explored in detail in an upcoming publication focused on $r$-process calculations with BSkG3.

\begin{figure}[htbp]
\begin{center}
\includegraphics[width=\linewidth]{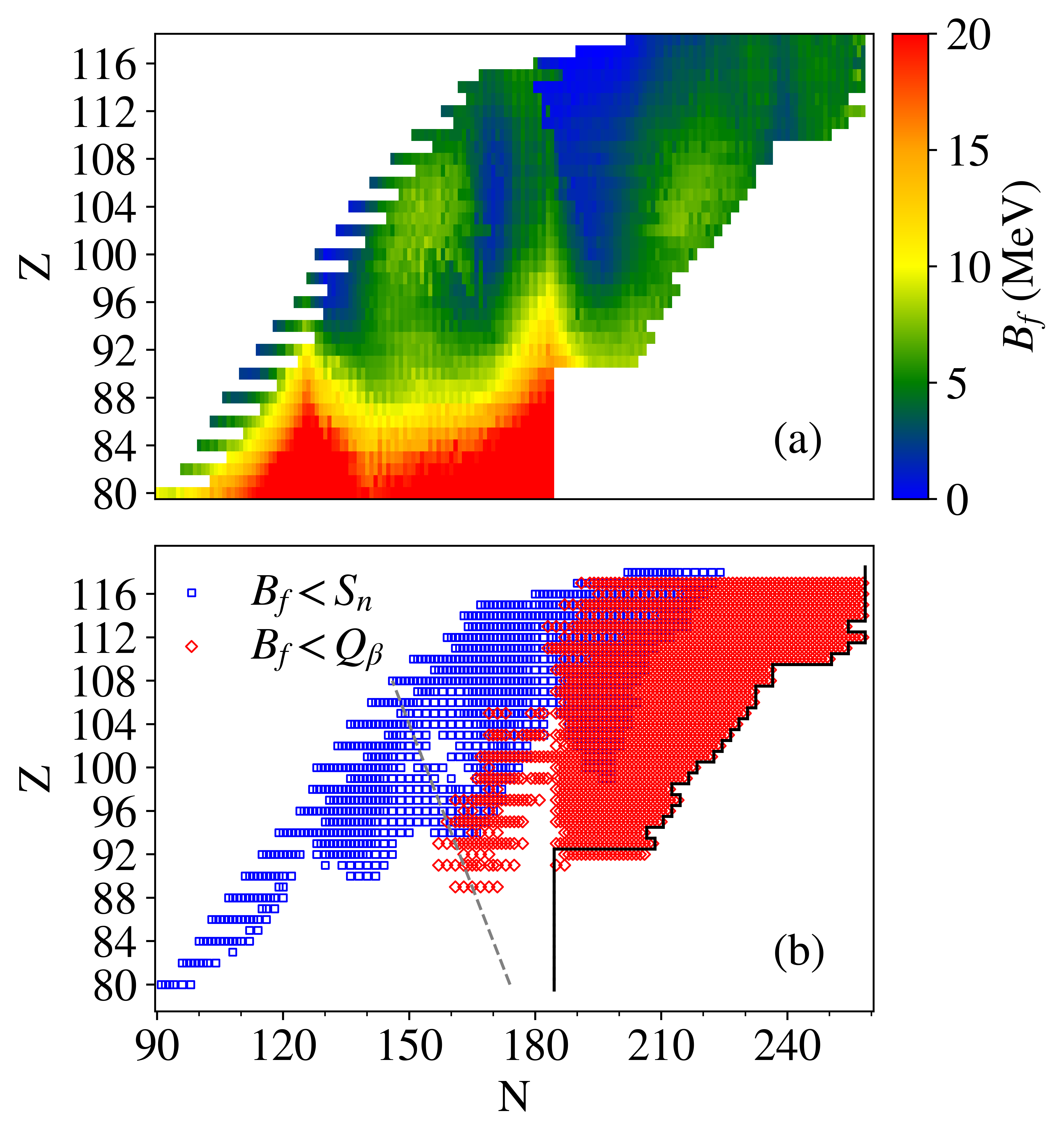}
\end{center}
\caption{\label{bf-sn} (a) Primary fission barriers, $B_f$, and (b) nuclei for which $B_f$ is lower than both $S_n$ and $Q_\beta$. The grey dashed line indicates the isobaric chain $A = 254$, while the black solid line marks the neutron drip line. All quantities are computed using the BSkG3 model.}
\end{figure}

The primary fission barriers also play a key role in assessing the impact of neutron-induced fission during the $r$-process, as they directly affect fission recycling~\cite{Goriely15}. In particular, neutron-induced fission is expected to suppress production via neutron capture when $B_f - S_n \sim 2$~MeV \cite{Panov2010}. In Fig.~\ref{bf-sn} (a), we present the full set of primary fission barriers in the ($N,Z$) plane. In addition, Fig.~\ref{bf-sn} (b) shows the nuclei whose primary barriers are smaller than $S_n$ and $Q_\beta$, thereby being likely to undergo n-induced or $\beta-$delayed fission. As anticipated in Fig.~\ref{comp_barriers}, we find that nuclei in the vicinity of $^{254}$Cf are highly susceptible to neutron-induced fission and more effectively their r-process isobaric progenitors to $\beta$-delayed fission, which inhibits the production of this nucleus after neutron irradiation. We also observe a region characterized by low $B_f$ is observed in the range $N \in [230,250]$ for $Z > 108$, in agreement with predictions from the BCPM and HFB-14 models. However, as we show in the next section, nuclei in this region will be susceptible to SF, which acts also as an effective barrier to the production of elements heavier than $Z\sim110$.

\section{Global overview of SF half-lives}\label{section_VI}

In the previous section, we showed that our predictions for the primary fission barrier are in relatively good agreement with other microscopic models across nearly the entire region $Z>80$, and even with ETFSI for nuclei with $90<Z<106$. However, agreement in SF half-lives is less straightforward, since deviations by 1~MeV in the primary fission barrier can impact SF half-lives by more than 5 orders of magnitude \cite{Sanchez2026a,Sanchez2026b}.

\subsection{Completing the model: collective inertias and path finding schemes}

\begin{figure*}[htbp]
\begin{center}
\includegraphics[width=\textwidth]{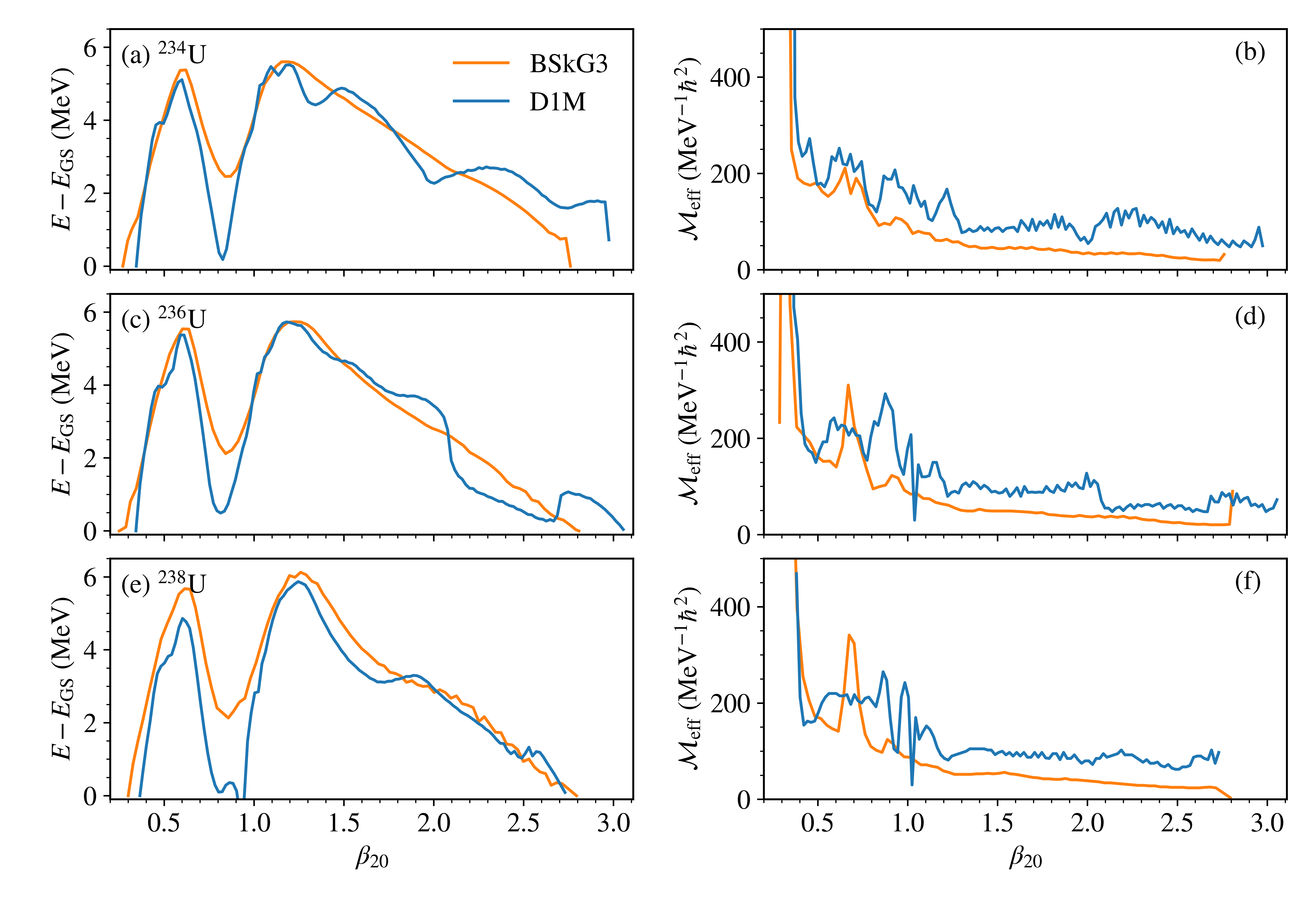}
\end{center}
\caption{\label{U_bskg_d1m}  Fission paths (left column) and effective inertias (right column) for $^{234}$U, $^{236}$U, and $^{238}$U obtained with BSkG3 (orange line) and D1M (blue line).}
\end{figure*}

Figure~\ref{U_bskg_d1m} compares the energy profiles and collective inertias along the fission paths of $^{234,236,238}$U obtained with the BSkG3 and D1M predictions. In both cases, the inertia tensor is computed within the ATDHFB framework. The one-dimensional fission paths predicted by the two functionals exhibit comparable barrier heights and similar path lengths. By contrast, although the collective inertias display analogous qualitative trends, the D1M functional yields systematically larger values, particularly beyond the outer barrier for $\beta_{20} \gtrsim 1.5$. This enhancement leads to larger collective actions and, consequently, significantly longer predicted half-lives. Specifically, within D1M the calculated half-lives are $4.8 \!\times\! 10^{39}$, $1.5 \!\times\! 10^{34}$, and $1.0 \!\times\! 10^{28}$ s for $^{234}$U, $^{236}$U, and $^{238}$U, respectively, whereas the corresponding BSkG3 predictions are $4.4 \!\times\! 10^{24}$, $5.7 \!\times\! 10^{24}$, and $9.8 \!\times\! 10^{24}$ s. This corresponds to deviations of more than 10 orders of magnitude, even though at first glance both models yield similar barrier heights and collective inertias.

\begin{figure*}[htbp]
\begin{center}
\includegraphics[width=\textwidth]{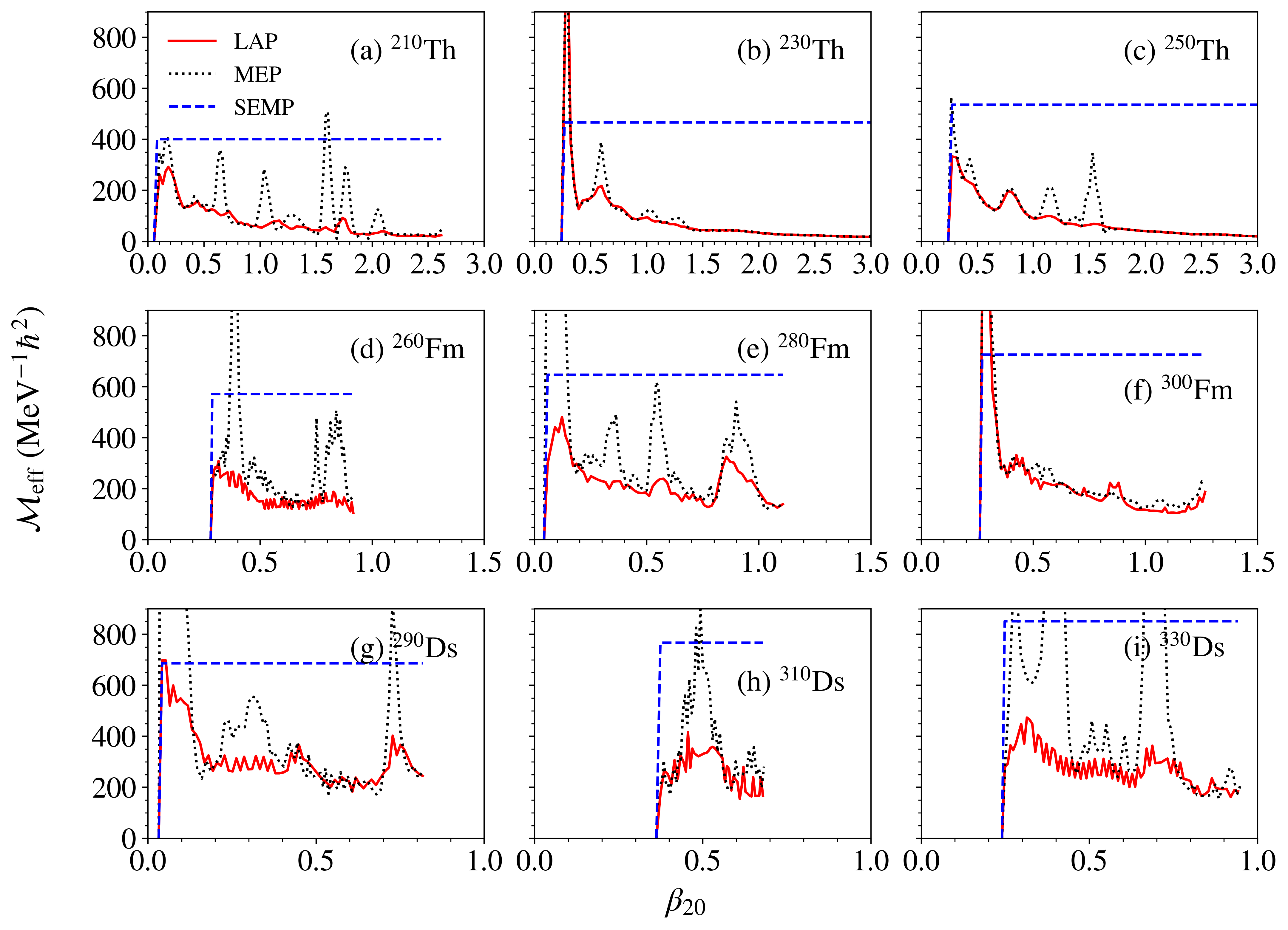}
\end{center}
\caption{\label{inertia_comp} Comparison between effective inertias along the LAP, MEP and the semiempirical value (Eq.~\ref{eq:semp_formula}) for selected nuclei.}
\end{figure*}

Even within the same nuclear model, the choice between minimizing the energy or the action integral can result in notable variations in the description of SF. In Fig. \ref{inertia_comp} we show for selected nuclei the collective inertia along the one-dimensional fission path when the action or the energy is minimized. In addition, we included the semiempirical effective inertia\cite{Nilsson1969}, commonly used to estimate fission cross sections \cite{capote2009,Goriely2009b,Koning23}:
\begin{equation}\label{eq:semp_formula}
    \mathcal{M}_{\rm eff}^{\rm SEMP} = 0.054 A^{5/3} \ \ \text{MeV} ^{-1}\hbar^2
\end{equation}
where $A$ is the mass number.

The semiempirical formula systematically produces large deformation-independent effective inertias compared to those obtained from the microscopic inertia tensor. The microscopic inertias display a pronounced dependence on the quadrupole deformation, generally decreasing as $\beta_{20}$ increases. Along fission paths that minimize the energy, we observe numerous peaks in the inertia (see the last row of Fig. \ref{inertia_comp}), arising from abrupt variations in $\beta_{22}$ and thus ultimately related to shell effects. Such a ``spiky'' behaviour is particularly important for describing SF in heavier nuclei, where, as illustrated in Fig. \ref{def_barr}, triaxiality remains significant even beyond the inner barriers. When the action integral is minimized, these peaks are substantially smoothed, leading to paths that preserve axiality, reduce the action, and consequently yield shorter SF half-lives. For instance, in the heaviest nucleus, $^{330}$Ds, the SF half-life decreases from 79.16 s to $2.8\!\times\!10^{-5}$ s. In contrast, using the semiempirical inertia results in a much longer half-life of $3.5\!\times\!10^{7}$ s. In fact, our calculations likely underestimate the influence of shell effects on the inertias due to the cranking approximation~\cite{Giuliani18}.

\subsection{Comparison with other models}
\label{sec:compother}

\begin{figure}[htbp]
\begin{center}
\includegraphics[width=\linewidth]{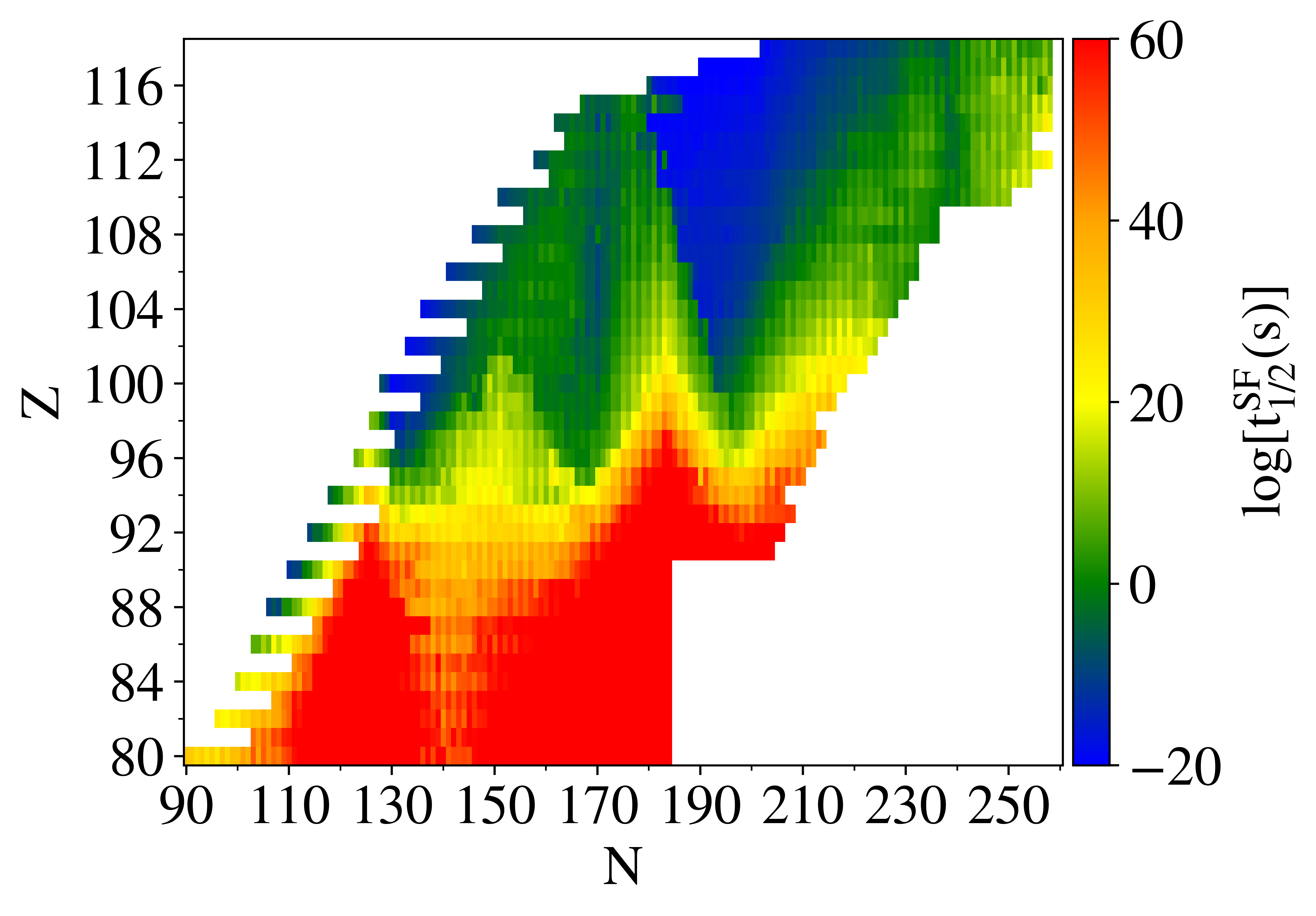}
\end{center}
\caption{\label{hf_BSkG3} Logarithm of the SF half-lives obtained with BSkG3.}
\end{figure}

\begin{figure}[htbp]
\begin{center}
\includegraphics[width=\linewidth]{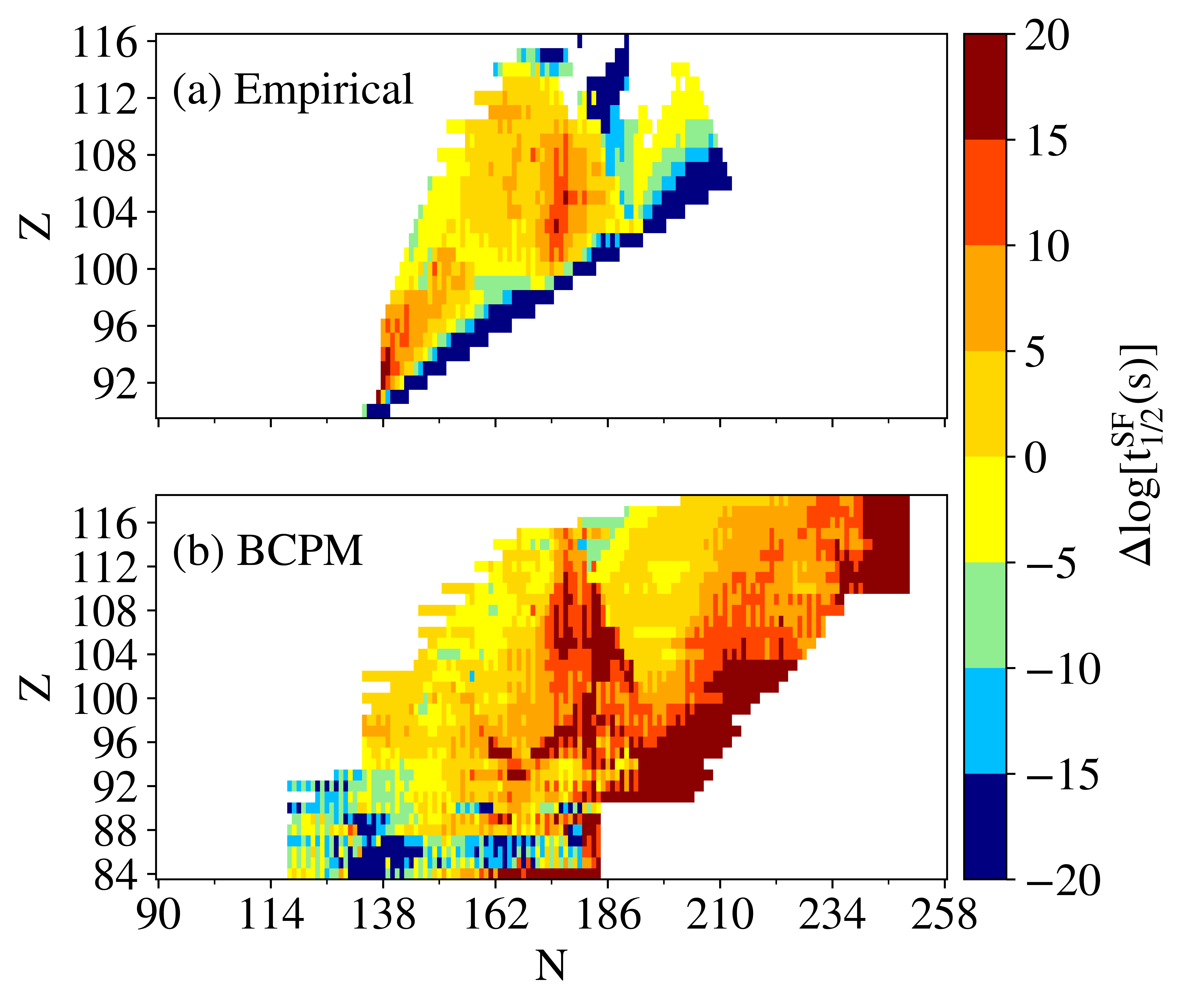}
\end{center}
\caption{\label{hf_comp} Difference in the decimal logarithm of the SF half-lives $t_{1/2}^{SF}$ between BSkG3 and (a) the values obtained with empirical formula from \cite{Khuyag20} using ETFSI primary barriers and (b) BCPM.}
\end{figure}

In Fig.~\ref{hf_BSkG3}, we present our SF half-lives covering the full region $80\leq Z\leq118$. Fig.~\ref{hf_comp}(a) shows the difference between our dataset and those obtained using the empirical formula of Ref.~\cite{Khuyag20} based on the ETFSI fission barriers. Finally, Fig.~\ref{hf_BSkG3}(b) shows the difference w.r.t. the BCPM predictions \cite{Giuliani18b}. In both figures, as usual, the decimal logarithm is used.

As shown in Fig.~\ref{hf_comp}(a), the SF half-lives obtained with BSkG3 are generally in agreement with those predicted by the empirical formula based on ETFSI barriers, with differences typically within 5--10 orders of magnitude. However, substantial discrepancies emerge toward the neutron-rich ends of the isotopic chains, where the empirical formula predicts extremely long half-lives. This is particularly noteworthy given that, as shown in Fig.~\ref{comp_barriers}, the primary fission barriers obtained with BSkG3 and ETFSI are generally comparable throughout the region considered. The large discrepancies in the neutron-rich region therefore suggest that, despite its reasonable performance in the experimentally constrained region, the empirical formula may not provide reliable extrapolations toward more neutron-rich nuclei. In particular, when applied to the heaviest isotopes in each chain, the formulas presented in Ref.~\cite{Khuyag20} yield negative values of the curvature energy, $\hbar\omega$, which characterizes the barrier width. Such values are non-physical, as they are incompatible with the expected positive curvature of the fission barrier at its saddle point.

 The BCPM set of SF half-lives was obtained using the microscopic inertia tensor derived from ATDHFB and, contrary to our approach, including a zero-point energy correction of 0.5 MeV in the action integral \cite{Giuliani18b}. Nuclei near the neutron drip line exhibit significantly longer half-lives with BSkG3 than those obtained with BCPM. A similar behaviour is found near the magic number $N=184$. This behavior is primarily attributed to the consideration of triaxially deformed GSs in BSkG3, which results in longer fission paths compared with those starting from the axial minimum. Again, in the region around $Z=108$ and $N\in[220,230]$, BSkG3 predicts half-lives of the order of seconds, making them competitive with neutron-capture rates and potentially acting as a barrier for the $r$-process nuclear flow. In this region, the half-lives predicted by BSkG3 are 10–15 orders of magnitude larger than those obtained with BCPM.  The actual impact on the $r-$process will be addressed in a forthcoming publication.

\section{Conclusions and outlook}\label{section_VII}

We performed calculations with the BSkG3 Skyrme-type EDF for nearly 3300 nuclei with $80<Z<118$, spanning the entire range between predicted proton and neutron drip lines. To do so, we computed more than 6000 PESs and deduced an equal number of fission paths that account for two quadrupole moments ($\beta_{20}, \beta_{22}$) and an octupole moment ($\beta_{30}$) as collective degrees of freedom. These paths are least-action paths, i.e.~that connect the GS to either a semiclassical turning point or a scission configuration in a way that minimizes the (classical) tunneling action calculated consistently from the PES and the associated microscopic ATDHFB inertia tensor evaluated with the cranking approximation. From these paths, we extracted the deformation and height of the fission barriers as well as the SF half-lives.

Similar to previous large-scale studies, our PESs are diverse in structure; the `double-hump` picture with an inner and outer barrier only applies to a subset of nuclei, among which the actinides with extracted empirical barriers. Many nuclei do not fit this traditional image: they exhibit just one barrier - e.g.~two zones of superheavy elements, ($Z \geq 104$, $N \geq 184$) and ($Z \geq 100$ and $N \leq 168 $), three barriers - e.g. the light actinides, where experimental evidence hints at the existence of a second isomer and third barrier~\cite{Blons1975,Schmidt2024} - or even four or more barriers. This diversity extends to the optimal nuclear shapes along the fission path. First, left-right asymmetry impacts GSs, primary and secondary barriers in different ways in several regions of the nuclear chart. Second - and for the first time in a study of this scale - we document that triaxiality impacts GSs, inner barriers and outer barriers as well; its influence is particularly strong for the inner barrier in the actinide region but it also significantly lowers the (single) barrier for several superheavy elements, in agreement with earlier, more limited studies~\cite{cwiok1996a,kowal2010,Abusara12}.

Our results for barrier heights \emph{qualitatively} match many features of earlier large-scale studies with D1M, BCPM and HFB-14. Nevertheless, there are significant quantitative differences that can be related to our more general treatment of deformation, and in particular the fact that we allow for triaxial deformation. The trends with $N$ and $Z$ of our calculated SF half-lives also match those of the earlier BCPM study~\cite{Giuliani18}, but we systematically predict that SF is slower for nearly all heavy nuclei.

We provide all of these results - together with saddle point nuclear level densities - in the current version of the TALYS reaction code and is also available at the BRUSLIB repository. As future work, we intend to generate a full set of neutron-induced and $\beta$-delayed fission rates intended for $r$-process simulations. At that point, the only remaining step toward a complete
and consistent set of fission data based on BSkG3 will be the prediction of fission fragment distributions and neutron multiplicities. Work in this direction based on time-dependent GCM simulations with FELIX is in progress~\cite{Regnier2018}.

Aside from our quest for completeness and the possibility of building EDFs that better grasp the physics of fission, we note that our results are not without issues. Although we have moved beyond collective spaces limited to a single dimension, the further inclusion of relevant collective coordinates such as the particle number dispersion can result in different LAPs, thereby changing the fission barriers we deduce and impact -- potentially dramatically --  the predicted SF half-lives. Addressing this challenge will require PESs constructed in collective spaces of three or more dimensions. This should be accompanied by a robust path-finding solver that can be generalized to arbitrary dimensions, at least if we continue to rely on the semiclassical WKB approximation. Generalising to a truly quantum-mechanical approach is also desirable~\cite{kawano2024}, even if this would require significant effort to generalize current reaction codes to be able to treat multi-dimensional PESs.

A more fundamental issue is our description of odd-mass and odd-odd systems. Although our PESs and collective inertias include the blocking effects of the odd particle(s), whether our collective modelling of fission  is appropriate for odd-mass and odd-odd systems is questionable~\cite{Bender20}. The fundamental issue is that these systems have high level density at low excitation energies, such that it is unlikely that their evolution from GS to scission is adiabatic. Unfortunately, it currently seems that a fully satisfactory solution might only be achieved through time-dependent simulations~\cite{bulgac2025a}. One smaller aspect of this issue might be solvable in the near future: an improved treatment of the collective inertias. Our treatment of the inertia relies on the cranking approximation, which is ill-defined for odd-mass and odd-odd nuclei. Linear response calculations with the finite-amplitude-method would give us access to the full, i.e.~non-approximate, ATDHFB inertias at a non-unsurmountable computational cost even in our 3D representation~\cite{Washiyama2021}. Contrary to energy differences along a PES, ATDHFB inertias might be sensitive to time-odd terms of the EDF and impact LAPs, barriers, SF half-lives and fragment distributions~\cite{ryssens2022,Bender20}. Such a refinement would even impact even-even systems, as the cranking approximation artificially suppresses shell effects compared to a complete treatment of the inertia tensor~\cite{Giuliani18}.

These proposed developments are but examples of a larger aspect: the future of nuclear fission depends on scientists successfully leveraging current and future computational resources. Such efforts should encompass both (i) a push to further improve algorithms for constrained nonlinear optimisation to tackle the HFB equations~\cite{baran2008,staszczak2010,ryssens2019} and (ii) the exploitation of novel hardware architectures (GPUs, APUs, \ldots). Particularly promising is the rise in different emulation techniques in different areas of nuclear theory~\cite{giuliani2023a,duguet2024,jin2025a}. Such techniques have the potential to dramatically lower the computational burden of large-scale surveys; emulation based on machine-learning has so far been deployed once~\cite{Lay24} while reduced-order-modelling techniques have, to the best of our knowledge, not found their way to nuclear fission so far. These techniques might be sufficiently powerful to enable the quantification of parametric uncertainty of fission properties at scale~\cite{schunck2015a}.

\appendix

\section*{Acknowledgments}

We are grateful to Drs.~L.~Robledo and S.~Giuliani for several enlightening discussions on the concept of collective inertia.
This work was supported by the Fonds de la Recherche Scientifique (F.R.S.-FNRS) and the Fonds Wetenschappelijk Onderzoek--Vlaanderen (FWO) under the EOS Projects No.~O000422 and O022818F, as well as by the F.R.S.-FNRS under the MIS Project No.~40028446. This research benefited from computational resources made available on the Tier-1 supercomputer Lucia of the Fédération Wallonie--Bruxelles, infrastructure funded by the Walloon Region under Grant Agreement No.~1117545. Additional computational resources were provided by the Consortium des Équipements de Calcul Intensif (CÉCI), funded by F.R.S.-FNRS under Grant No.~2.5020.11 and by the Walloon Region. Additionally, resources and services used in this work were provided by the VSC (Flemish Supercomputer Center), funded by the Research Foundation – Flanders (FWO) and the Flemish Government. We acknowledge the EuroHPC Joint Undertaking for awarding access to the MareNostrum5 supercomputer at the Barcelona Supercomputing Center (BSC), Spain, under projects EHPC-DEV-2025D04-097 and EHPC-DEV-2026D01-043. We further acknowledge EPICURE, a EuroHPC Joint Undertaking initiative, for supporting our work on MareNostrum5 in the context of these development projects.

\section{Structure of the fission path files}


The $(\beta_{20}, \beta_{22})$ LAPs, as compiled in TALYS and BRUSLIB, are organized by isotopic chain covering nuclei from Hg ($Z=80$) to Og ($Z=118$), formatted for direct use with the TALYS code. Each LAP begins with a header specifying $Z$, $A$, $\texttt{nbeta}$ (number of points along the path), and $\texttt{Egs}$ (GS energy determined on the $(\beta_{20}, \beta_{22})$ potential-energy surface). The header is followed by a table of $\texttt{nbeta}$ rows describing the evolution along the path. Each row contains the columns $\texttt{beta20}$, $\texttt{beta22}$, $\texttt{beta3}$, $\texttt{E-Egs}$, $\texttt{MI\_ATD}$, and $\texttt{IP}$, corresponding respectively to the quadrupole deformation $\beta_{20}$, triaxial deformation $\beta_{22}$, octupole deformation $\beta_{30}$, the effective potential $V_{\rm eff}$ measured relative to the GS, the collective inertia $\mathcal{M}_{\rm eff}$, and an integer flag identifying special points along the path.

The $\texttt{IP}$ flag distinguishes key configurations along the LAP: the entrance point $s_{\rm in}$ is assigned $\texttt{-1}$, local maxima are assigned $\texttt{1}$, and local minima are assigned $\texttt{2}$. If more than three local maxima are present, only the three highest are labeled $\texttt{1}$, while the remaining maxima are labeled $\texttt{8}$. Similarly, only the two minima located between the three highest maxima are assigned $\texttt{2}$, while any additional minima are labeled $\texttt{9}$.

The effective inertia in the files is defined with respect to a dimensionless path parameter $s$ with uniform spacing, $ds = 1$. Therefore, the action integral, and hence the SF half-lives, can be calculated directly from the path files as

\begin{equation}
S\approx\sum_{i=1}^{N_{\rm path}}\sqrt{2\mathcal{M_{\rm eff,i}}E_i}
\end{equation}
where $N_{\rm path}$ is the total number of points along the LAP. 

To express the inertia in terms of the deformation coordinate $\beta_{20}$, a change of variables is required,
\begin{equation}
    \mu(\beta_{20}) = \mathcal{M}_{\rm eff}(s)\left(\frac{ds}{d\beta_{20}}\right)^2 \,,
\end{equation}
which, in the discrete representation of the path, can be approximated as
\begin{equation}
    \mu(\beta_{20}) \approx \mathcal{M}_{\rm eff}(s)\,\frac{1}{(\Delta \beta_{20})^2}\,.
\end{equation}
Additional files, including a summary table of saddle points, potential wells, and SF half-lives, are available in the BRUSLIB repository~\cite{Bruslib}.

\bibliography{apssamp}

\end{document}